\documentclass[twocolumn,aps,showpacs,raggedbottom,nobalancelastpage,amssymb,superscriptaddress]{revtex4-1}

\usepackage{epsfig,amssymb,amsmath,latexsym}
\usepackage{subfigure}
\usepackage{wrapfig}
\usepackage{float}
\usepackage{dsfont}
\usepackage{accents}
\usepackage{physics}
\usepackage{tensor}
\usepackage{graphicx}
\usepackage{esint}
\newcommand\bigzero{\makebox(0,0){\text{\huge0}}}

\usepackage[dvipsnames]{xcolor}
\usepackage{soul}
\setulcolor{red}
\setstcolor{red}
\sethlcolor{yellow}

\begin{document}
\title{Pumped heat and charge statistics from Majorana braiding}
\author{Thomas Simons}
\affiliation{Department of Physics, Lancaster University, Lancaster LA1 4YB, United Kingdom}
\author{Dganit Meidan}
	\affiliation{Department of Physics, Ben-Gurion University of the Negev, Beer-Sheva 84105, Israel} 
\author{Alessandro Romito}
\affiliation{Department of Physics, Lancaster University, Lancaster LA1 4YB, United Kingdom}

\date{\today}
\begin{abstract}
We examine the heat and charge transport of a driven topological superconductor. Our particular system of interest consists of a Y-junction of topological superconducting wires, hosting non-Abelian Majorana zero modes at their edges. The system is contacted to two leads which act as continuous detectors of the system state. We calculate, via a scattering matrix approach, the full counting statistics of the driven heat transport, between two terminals contacted to the system, for small adiabatic driving and characterise the energy transport properties as a function of the system parameters (driving frequency, temperature). We find that the geometric, dynamic contribution to the pumped heat statistics results in a correction to the Gallavotti-Cohen type fluctuation theorem for quantum heat transfer. Notably, the correction term to the fluctuation theorem extends to cycles which correspond to topologically protected braiding of the Majorana zero modes. This geometric correction to the fluctuation theorem differs from its analogs in previously studied systems in that (i) it is non-vanishing for adiabatic cycles of the system's parameters, without the need for cyclic driving of the leads and (ii) it is insensitive to small, slow fluctuations of the driving parameters due to the topological protection of the braiding operation.
  
\end{abstract}

%\hskip 3cm\begin{minipage}[l]{12cm}
\maketitle
%\end{minipage}
%%%%%%%%%%%%%%%%%%%%%%%%%%%%%%%%%%%%%%%%%%%%%%%%%%%%%%%%%%%%%%%%%%
\section{Introduction}
\label{intro}
The time dependent, cyclic evolution of the internal parameters of a quantum system can lead to the accumulation of a geometric phase that depends solely on the path traversed in parameter space and not on the duration of the cycle itself \cite{Berry1984a}. This geometric phase manifests itself in different phenomena in all areas of physics \cite{cohen2019geometric}.
Geometric contributions to quantum evolution become particularly interesting for many-body systems in a topological phase of matter, hosting non-Abelian excitations at their edges \cite{wilczek1990fractional}. In this case, the phase is generalized to a unitary operation protected against details of the system and the evolution, which makes anyons appealing excitations for potential use in fault tolerant quantum computation \cite{Nayak2008,Aasen2016}. A particularly interesting case of non-Abelian zero energy excitations are Majorana zero modes, which exist on the surface of topological superconductors \cite{read2000paired,kitaev2001unpaired}. The proposal to engineer topological superconductivity in semiconductor nanostructures \cite{oreg2010helical,sau2010generic,Alicea2012,lutchyn2018majorana} has received compelling experimental indications \cite{mourik2012signatures,deng2016majorana,albrecht2016exponential,zhang2018quantized,lutchyn2018majorana} and led to proposals for quantum information processing \cite{Aasen2016,Sarma2015,Plugge2016,Karzig2017}. 

The experimental indications of Majorana zero modes are based on charge transport measurements, reflecting the identification of topological indices in scattering processes for topological superconductors \cite{akhmerov2011quantized,fulga2012scattering,meidan2014scattering}. While these experiments are sensitive to the existence of protected zero modes, their charge neutrality makes a direct manifestation of the topologically protected unitary evolution more elusive to detection via charge transport. Heat transport does not suffer from this limitation. It appears to be non-trivially affected by Majorana zero modes \cite{smirnov2018universal,smirnov2019dynamic,ricco2018tuning,molignini2017sensing}, and geometric effects of Majorana braiding have been  predicted for heat pumped through braiding \cite{Meidan2019}. 

Geometric contributions have generically found to be evident in transport processes such as pumped charge and heat currents in cyclically manipulated quantum systems with few degrees of freedom \cite{Ren2010,Placke2018,Yuge2012,KumarYadalam2016,Makhlin2001,Bhandari2020,Takahashi2020,Andreev2000,Brouwer1998,Shutenko2000,Avron2001,Moskalets2002,bhandari2020geometric}. For these systems with few degrees of freedom, recent studies \cite{Ren2010,Agarwalla2012,Goswami2016a} have suggested that geometric contributions to the full statistics of transfer processes result in the apparent violation of fluctuation theorems, which quantify the likelihood of anomalous heat transfer against a thermal gradient \cite{Gallavotti1995}.

Motivated by these findings, it becomes of interest to investigate how the topological protection, of the geometric phases in Majorana based manipulations, is reflected in the corrections of the aforementioned fluctuation theorem. In this spirit, we further explore the influence of geometric contributions to the full counting statistics of pumped heat transport, in the case of the exchange of two Majorana fermions performed within a Y-junction of topologically superconducting nanowires. We address our interest specifically to the effect upon fluctuation theorems. By using a scattering matrix approach, we will show that we find a non-zero geometric contribution to the probability generating function and that this contribution does indeed lead to a correction to the Gallavotti-Cohen type fluctuation theorem. Such a correction generically exisit for  arbitrary adiabatic cycles in parameter space, but it extends to the case of Majorana braiding, in which it becomes insensitive to slow time-fluctuations of the driving parameters.

The paper is organized as follows. In Sec. \ref{sec:FCS} we develop the general formalism to compute the full counting statistics of energy transfer via scattering matrices, including both particle and hole degrees of freedom required for superconducting systems. We then address the protocols of interest in Sec. \ref{Braiding Sec} , where we analyze the scattering matrix for a driven Y-junction of 1-dimensional p-wave superconductors.  We employ our formalism to compute the transport properties beyond the average current in Sec. \ref{sec:transport} and the corrections to the Gallavotti-Cohen fluctuation theorem in Sec. \ref{sec:theorem}. In the latter we first address pumping cycles of small amplitude and finally extend our results to topologically protected braiding, where we characterize the topological features in the corrections to the mentioned fluctuation theorem.

%%%%%%%%%%%%%%%%%%%%%%%%%%%%%%%%%%%%%%%%%%%%%%%%%%%%%%

%%%%%%%%%%%%%%%%%%%%%%%%%%%%%%%%%%%%%%%%%%%%%%%%%%%%%%

\section{Full Counting Statistics for Pumped Heat Transport}
\label{sec:FCS}
In order to study the behaviour of thermal fluctuations throughout any pumped process, it is necessary to extract information beyond the average pumped quantities and hence uncover the full statistical distribution of the transport process. Such information is provided by the probability distribution, $P(Q,\mathcal{T})$, for some quantity of interest $Q$, e.g.\ charge or energy, transported across a system throughout some time period $\mathcal{T}$. This distribution can be accessed via its Fourier transform, the characteristic function (CF) $\chi(\lambda)$, where $\lambda$ denotes the counting field. The full counting statistics of charge transfer, originally introduced for DC transport \cite{levitov1996electron},  has previously been evaluated for pumped electronic charge \cite{Makhlin2001,Andreev2000} and for specific non-adiabatic periodic driving of superconducting devices \cite{romito2004full}.  For the case of a Majorana braiding, for which topological features are apparent in scattering properties, we construct the FCS  based on the scattering matrix formalism. We consider a superconducting system under the cyclic modulation of some internal parameters, which in this case correspond to the couplings between the external and central Majorana states present in a superconducting Y-junction (cf.\ Fig.\ \ref{Setup}). This time dependent manipulation facilitates inelastic scattering events and as a result, it is important to carefully consider both the energy and time dependence of the scattering events when defining the CF. We define the CF for the heat, $Q$, pumped during the total cycle period $\mathcal{T}$ as
\begin{equation}
\label{Total Char Func 1}
\chi_Q(\lambda) = \int dQ e^{i\lambda Q} P(Q,\mathcal{T}).
\end{equation}  
The probability distribution for the total cycle, $P(Q,\mathcal{T})$, can be obtained by considering the heat transported during the time steps, $t_i$, of a discretised cycle: 
\begin{equation}
\label{Total Char Func}
\chi_Q(\lambda) = \int dQ e^{i\lambda Q} \sum_{\{q_{t_i}\}} \delta \Big (Q=\sum_{t_i}q_{t_i} \Big ) P(q_{t_1},q_{t_2},...),
\end{equation}       
where $\{q_{t_i}\}$ denotes all possible combinations of the heat quantities $q_{t_i}$, transported in each discrete time step $t_i$. 

By considering the particle baths in the external leads to be large, so that the ingoing distribution function at any time $t$ is given by the equilibrium Fermi distribution function, and if relaxation times are fast enough, we can assume independence of the probability distribution at each time. The CF can then be written as a product of the contribution from each time step:
\begin{equation}
\label{Total Char Func Int}
\begin{aligned}
\chi_Q(\lambda) =& \sum_Q e^{i\lambda Q} \sum_{\{q_{t_i}\}} \delta \Big (Q=\sum_{t_i}q_{t_i} \Big ) \prod_{t_i} P(q_{t_i}) 
\\
=&  \sum_{\{q_{t_i}\}}  e^{i\lambda \sum_{t_i} q_{t_i}} \prod_{t_i} P(q_{t_i}) 
\\
=& \sum_{q_{t_1}} e^{i\lambda  q_{t_1}} \sum_{q_{t_2}}e^{i\lambda  q_{t_2}} ... \prod_{t_i} P(q_{t_i})  
\\
=& \prod_{t_i} \sum_{q_{t_i}} e^{i\lambda  q_{t_i}}P(q_{t_i})  = \prod_{t_i} \chi_{t_i}(\lambda).
\end{aligned}
\end{equation}
Taking the continuous limit, $t_i \to 0$, we can write the cumulant generating function (CGF), defined as $G_Q(\lambda) = \ln(\chi_Q(\lambda))$ as an integral over the driving time period:
\begin{equation}
\label{Generating Function}
G_Q(\lambda) = \int_0^\mathcal{T} dt \ln(\chi_t(\lambda)).
\end{equation}
We have therefore reduced the calculation, of the total FCS of driven heat transport, to that of a CF at a frozen time $t$, which we denote as $\chi_t(\lambda)$.  The latter is computed analogously to the case of charge FCS \cite{Muzykantskii1994b}, as outlined in Appendix \ref{CF derivation}: 
\begin{equation}
\label{Char Func t}
\chi_t(\lambda) =  \Big < e^{i\lambda \hat{Q}_\rightarrow(t)}e^{-i\lambda \hat{Q}_\leftarrow(t)} \Big >.
\end{equation}
Here the operators $\hat{Q}_{\rightarrow(\leftarrow)}(t)$ describe the heat carried by particles in the left lead, entering (leaving) the junction with the internal system of interest, at some time $t$. The CF in Eq. \ref{Char Func t} has been previously evaluated for the case of charge transfer between superconducting leads in static systems by B.A. Muzykantskii and D.E. Khmelnitskii \cite{Muzykantskii1994b}. We extend, hereafter, this formalism to the case of heat transport and adiabatically driven systems.

The heat operators can be written in terms of fermionic creation and annihilation operators in the left external lead \cite{Moskalets2014}: 
\begin{equation}
\begin{aligned}
\label{Ingoing Heat}
\hat{Q}_{\rightarrow}(t) = \iint_{-\infty}^\infty d\epsilon d\epsilon' & \Big(\frac{\epsilon+\epsilon'}{2} \Big)  e^{\frac{i(\epsilon-\epsilon')t}{\hbar}}
\\
& \times \Big(\hat{a}_{L^e}^\dagger(\epsilon) \hat{a}_{L^e}(\epsilon') +\hat{a}_{L^h}^\dagger(\epsilon) \hat{a}_{L^h}(\epsilon')\Big)
\\
\end{aligned}
\end{equation}
and
\begin{equation}
\begin{aligned}
\label{Outgoing Heat}
\hat{Q}_{\leftarrow}(t) = \iint_{-\infty}^\infty d\epsilon d\epsilon' & \Big(\frac{\epsilon+\epsilon'}{2} \Big)  e^{\frac{i(\epsilon-\epsilon')t}{\hbar}}
\\
& \times \Big(\hat{\phi}_{L^e}^\dagger(\epsilon) \hat{\phi}_{L^e}(\epsilon') +\hat{\phi}_{L^h}^\dagger(\epsilon) \hat{\phi}_{L^h}(\epsilon')\Big).
\end{aligned}
\end{equation}
The ingoing, $\hat{a_{i}}$, and outgoing, $\hat{\phi_{i}}$, electron (e) and hole (h) scattering states in the left (L) and right (R) leads are related by the scattering matrix:
\begin{eqnarray}
\label{scattering relation}
\begin{pmatrix}
\hat{\phi}_{L^e}(\tilde{\epsilon})
\\
\hat{\phi}_{L^h}(\tilde{\epsilon})
\\
\hat{\phi}_{R^e}(\tilde{\epsilon})
\\
\hat{\phi}_{R^h}(\tilde{\epsilon}) 
\end{pmatrix}
= S_F(\tilde{\epsilon},\epsilon)
\begin{pmatrix}
\hat{a}_{L^e}(\epsilon)
\\
\hat{a}_{L^h}(\epsilon)
\\
\hat{a}_{R^e}(\epsilon)
\\
\hat{a}_{R^h}(\epsilon)
\end{pmatrix},
\end{eqnarray}
where the scatering matrix depends explicitly on two energies due to the fact that the scattered particles can absorb or emit energy, due to the external driving. For a periodically driven system, as we are considering here, energy can be absorbed or emitted only in multiples of the driving frequency, so that $\tilde{\epsilon}=\epsilon-n\omega$. With this relationship between ingoing and outgoing states, Eq.\ \ref{Char Func t}, \ref{Ingoing Heat} and \ref{Outgoing Heat} allow the FCS for heat transport to be accessed entirely via the full Floquet scattering matrix describing transport across the internal system.   

%%%%%%%%%%%%%%%%%%%%%%%%%%%%%%%%%%%%%%%%%%%%%%%%%%%%%%%%%%%%%%%%%%%%%%%%%%%%%%%%%%%%%%%%%%%%%%%%%%%%%%%%%%%%%%%%%%%%%%%%%%%%%%%%%%%%%%%%%%%%%%%%%%%%%%%%%%%%%%%%%%%%%%%%%%%%%%%%%%
\subsection{Adiabatic and Small Driving Amplitude Limit}
\label{sec:Adiabatic}
In general, for a time dependent driven system it is difficult to determine the elements of the full Floquet scattering matrix, which accommodates for all possible inelastic scattering events induced by the driving. In order to make analytical progress, we choose to study a model subjected to two important approximations. Firstly, we assume that the periodic driving of the system is adiabatic, in the sense that the driving period, $\mathcal{T}$, is large compared to the scattering time. In this situation, scattering can be considered instantaneous and described by a frozen scattering matrix, $S(\epsilon,t)$, the properties of which are modulated periodically by the driving. More precisely, if the driving is switched off, the Floquet Scattering matrix in Eq.\ (9), $S_F(\epsilon, \epsilon)$, describes an energy dependent, time-translation-invariant scattering process. If the matrix depends on time  via a parameter, one can consider such a frozen scattering matrix parametrically depending on time, $S_{F,t}(\epsilon, \epsilon) \equiv S(\epsilon,t)$. Secondly, we assume that the amplitude of the driving, in the relevant parameter space of the system, is small. In Sec. \ref{Large Amp Drive} we will show that the results yielded from this approach can be extended to the case of the Majorana braiding, for which the amplitude of the driving can no longer be consider small with respect to the values of the parameters at the center of the cycle. 

The weak, adiabatic, periodic driving of parameters, such as the lead temperatures or lead coupling strength, with frequency $\omega$ can be modeled as
\begin{equation}
\label{parameter modulation}
X_j(t) = X_{j,0} + X_{j,\omega}e^{i(\omega t-\eta_j)} + X_{j,\omega}e^{-i(\omega t-\eta_j)}. 
\end{equation} 
 With the assumption that the amplitude of this modulation, $X_{j,\omega}$, is small enough to expand to first order, then the corresponding time dependence of the scattering matrix can be expressed as \cite{Moskalets2002}
\begin{eqnarray}
\label{S expand}
\begin{aligned}
S(\epsilon,t) \approx& \ S(\epsilon,X_{j,0}) + S_\omega(\epsilon) e^{-i\omega t} +  S_{-\omega}(\epsilon) e^{i\omega t},
\\
\mathrm{where} \ S_{\pm \omega} =& \sum_j X_{j,\omega}e^{\mp i\eta_j} \frac{\partial S}{\partial X_j}.
\end{aligned}
\end{eqnarray}
The scattering matrix in this form corresponds to a zeroth order expansion in frequency of the full Floquet scattering matrix, whilst allowing only scattering processes between nearest energy sidebands in addition to elastic events. The corresponding operators for scattered states then take the form 
\begin{equation}
\begin{aligned}
{\hat{\phi}_i(\epsilon)} = \sum_{\alpha}\Big ( S^{i\alpha }(\epsilon){\hat{a}_\alpha(\epsilon)}& + {S_{-\omega}^{i\alpha }(\epsilon)}{\hat{a}_\alpha(\epsilon + \omega)}
\\
&+ {S_{+\omega}^{i\alpha }(\epsilon)}{\hat{a}_\alpha(\epsilon - \omega)} \Big ),
\end{aligned}
\end{equation}
where creation and annihilation operators for the four ingoing channels at energy $\epsilon_i$ are defined in Eq. \ref{scattering relation}.
Before using this approximation of the scattering matrix in the expressions for the ingoing and outgoing heat operators (Eqs. \ref{Ingoing Heat}, \ref{Outgoing Heat}), we notice that upon calculation of the CGF in Eq. \ref{Generating Function}, and hence integration over the driving time period, only terms where $\epsilon=\epsilon'$ will contribute to the heat operators. Consequently, the evaluation of the CGF only requires the calculation of the ingoing and outgoing number operators at a single energy, defined as
\begin{equation}
\begin{aligned}
& \hat{N}^{e(h)}_\rightarrow (\epsilon) = \hat{a}^\dagger_{L^{e(h)}} (\epsilon) \hat{a}_{L^{e(h)}} (\epsilon) 
 \\
\mathrm{and} & \ \hat{N}^{e(h)}_\leftarrow (\epsilon) = \hat{\phi}^\dagger_{L^{e(h)}} (\epsilon) \hat{\phi}_{L^{e(h)}} (\epsilon).
\end{aligned}
 \end{equation}
The number operators for both the ingoing and outgoing particle states can then be expressed in terms of matrices, $P$, acting on the ingoing scattering states:
\begin{equation}
\hat{N}^{e(h)}_{ \rightarrow(\leftarrow)}(\epsilon_l) = \sum_{\substack{\alpha,\beta   \\ \epsilon_i,\epsilon_j}} \Big [ P^{e(h)}_{\rightarrow(\leftarrow)}(\epsilon_l) \Big ]^{\alpha \beta}_{\epsilon_i\epsilon_j} {\hat{a}^\alpha(\epsilon_i)}^\dagger \hat{a}^\beta(\epsilon_j),
\end{equation}
with $\alpha, \beta \in \{L^e,L^h,R^e,R^h\}$. Here the ingoing scattering matrices are diagonal in both the discretised energy and the electron-hole bases:
\begin{equation}
\label{ingoing P}
\Big [ P^{e(h)}_{\epsilon_l \rightarrow} \Big ]^{\alpha \beta}_{\epsilon_i\epsilon_j} = \delta_{\alpha L^{e(h)}}\delta_{\alpha \beta}\delta_{\epsilon_i \epsilon_l}\delta_{\epsilon_i \epsilon_j}.
\end{equation}
However, the inelastic scattering events, induced by the driving, result in non-diagonal matrices defining the outgoing number operators:
\begin{widetext}
\begin{equation}
\label{outgoing P}
\begin{aligned}
\Big [ P^{e(h)}_{\epsilon_l \leftarrow} \Big ]^{\alpha \beta}_{\epsilon_i\epsilon_j} =& \delta_{\epsilon_i \epsilon_l}{S^{\alpha L^{e(h)}}(\epsilon_l)}^* \Big (  S^{L^{e(h)} \beta}(\epsilon_l) \delta_{\epsilon_i \epsilon_j} +  S^{L^{e(h)} \beta}_{-\omega}(\epsilon_l) \delta_{(\epsilon_i + \omega) \epsilon_j} +  S^{L^{e(h)} \beta}_{\omega}(\epsilon_l) \delta_{(\epsilon_i - \omega) \epsilon_j}  \Big )
\\
+& \delta_{\epsilon_i (\epsilon_l+\omega)}{S^{\alpha L^{e(h)}}_{-\omega}(\epsilon_l)}^* \Big (  S^{L^{e(h)} \beta}(\epsilon_l) \delta_{(\epsilon_i-\omega)\epsilon_j} +  S^{L^{e(h)} \beta}_{-\omega}(\epsilon_l) \delta_{\epsilon_i \epsilon_j} +  S^{L^{e(h)} \beta}_{\omega}(\epsilon_l) \delta_{(\epsilon_i - 2\omega) \epsilon_j}  \Big )
\\
+& \delta_{\epsilon_i (\epsilon_l-\omega)}{S^{\alpha L^{e(h)}}_{\omega}(\epsilon_l)}^* \Big (  S^{L^{e(h)}\beta}(\epsilon_l) \delta_{(\epsilon_i+\omega)\epsilon_j} +  S^{L^{e(h)} \beta}_{-\omega}(\epsilon_l) \delta_{(\epsilon_i + 2\omega) \epsilon_j} +  S^{L^{e(h)} \beta}_{\omega}(\epsilon_l) \delta_{\epsilon_i  \epsilon_j}  \Big ).
\end{aligned}
\end{equation}
\end{widetext}
Using these matrices $P$ in Eq. \ref{Char Func t}, the characteristic function can be expressed as the average of a product of exponentials:
\begin{equation}
\label{CF expectation}
\chi_t(\lambda) = \Big < \mathrm{exp} \Big (i \lambda \sum_{\alpha,\beta} C_{\alpha \beta} \hat{a}_\alpha^\dagger \hat{a}_\beta \Big )  \mathrm{exp} \Big (-i \lambda \sum_{\alpha,\beta} D_{\alpha \beta} \hat{a}_\alpha^\dagger \hat{a}_\beta \Big ) \Big >,
\end{equation}  
with $C = \sum_i \epsilon_i  P_{\epsilon_i \rightarrow}$ $D = \sum_i \epsilon_i  P_{\epsilon_i \leftarrow}$ and the sum of the electron and hole number operator matrices defined as $P_{\epsilon_i \rightarrow}=P^e_{\epsilon_i \rightarrow} + P^h_{\epsilon_i \rightarrow}$. The relevant density matrix, $\rho$, is block diagonal in the energy basis with the block at each energy $\epsilon_i$ being given by $\rho^{\epsilon_{l}}_{ij} = \langle \hat{a}^{i \dagger}_{\epsilon_{l}} \hat{a}^j_{\epsilon_{l}} \rangle = f_i(\epsilon_{l})\delta_{ij}$. Importantly $P$ are projective matrices, $P^2=P$, as shown in Appendix \ref{projector derivation}. Under this condition, it has been proven \cite{Muzykantskii1994b} that the expectation value in Eq. \ref{CF expectation} can be expressed as a determinant via
\begin{equation}
\begin{aligned}
\label{det form 1}
&\chi_{t}(\lambda) = \det (1-\rho+\rho e^{i\lambda C}e^{-i\lambda D}) 
\\
=& \det(1-\rho+\rho e^{i \lambda \sum_i \epsilon_i P_{\epsilon_{i}\rightarrow}} \Big(1 + \sum_i P_{\epsilon_{i}\leftarrow} (e^{-i \lambda \epsilon_{i}}-1) \Big))
\\
&= \det(M_t(\lambda)).
\end{aligned}
\end{equation}
In general the matrix $M_t(\lambda)$ will be of block pentadiagonal form in an infinite energy basis. In order to make analytical progress we can split the matrix $P_{\epsilon_{i}\leftarrow}$ into two contributions as $P_{\epsilon_i \leftarrow}^0 + \tilde{P}_{\epsilon_i \leftarrow}$, where $P_{\epsilon_i \leftarrow}^0$ describes the part of the matrix which survives in the static limit and $\tilde{P}_{\epsilon_i \leftarrow}$ includes all contributions that arise from the periodic driving and hence all terms involving the sideband scattering matrix coefficients $S_{\pm \omega}$. Subsequently we can split the matrix $M_t$ as $M_{t,0}+\tilde{M}_t$, where
\begin{widetext}
\begin{equation}
\label{GF Full Form}
\begin{aligned}
& M_{t,0} =  1 - \rho + \rho \exp(i \lambda \sum_i \epsilon_i P_{\epsilon_i \rightarrow})\Bigg(1 + \sum_i P_{\epsilon_i \leftarrow}^0 (e^{-i \lambda \epsilon_i}-1)\Bigg), 
\\
&\tilde{M}_t =   \rho \exp(i \lambda \sum_i \epsilon_i P_{\epsilon_i \rightarrow}) \Bigg ( \sum_i  \tilde{ P}_{\epsilon_i \leftarrow} (e^{-i \lambda \epsilon_i}-1) \Bigg ),
\end{aligned}
\end{equation}
\begin{equation}
\nonumber
\begin{aligned}
& P_{\epsilon_i \leftarrow}^{e,0} =  \begin{pmatrix}
\ddots
\\
& \bigzero
\\
& & 0 & 0 &0 & & 
\\
& &  0 & {S^{\alpha L^{e}}(\epsilon_i)}^*S^{L^{e} \beta}(\epsilon_i) & 0 & & 
\\
& & 0 & 0 & 0 & & 
\\
& & & & & \bigzero
\\
& & & & & & \ddots
\end{pmatrix}
\\
& \mathrm{and}
\\
&  \tilde{P}^e_{\epsilon_i \leftarrow} =  \begin{pmatrix}
\ddots
\\
& \bigzero
\\
& & {S_\omega^{\alpha L^e}(\epsilon_i)}^*S_\omega^{L^e \beta}(\epsilon_i) & {S_\omega^{\alpha L^e}(\epsilon_i)}^*S^{L^e \beta}(\epsilon_i) & {S_\omega^{\alpha L^e}(\epsilon_i)}^*S_{-\omega}^{L^e \beta}(\epsilon_i) & & 
\\
& & {S^{\alpha L^e}(\epsilon_i)}^*S_\omega^{L^e \beta}(\epsilon_i) & 0 & {S^{\alpha L^e}(\epsilon_i)}^*S_{-\omega}^{L^e \beta}(\epsilon_i) & & 
\\
& & {S_{-\omega}^{\alpha L^e}(\epsilon_i)}^*S_\omega^{L^e \beta}(\epsilon_i) & {S_{-\omega}^{\alpha L^e}(\epsilon_i)}^*S^{L^e \beta}(\epsilon_i) & {S_{-\omega}^{\alpha L^e}(\epsilon_i)}^*S_{-\omega}^{L^e \beta}(\epsilon_i) & & 
\\
& & & & & \bigzero
\\
& & & & & & \ddots
\end{pmatrix}.
\end{aligned}
\end{equation}
\end{widetext}
 With these definitions, the CGF, $G(\lambda) = \ln \chi(\lambda)$, can be expressed as a sum of two contributions:
\begin{equation}
\begin{aligned}
\label{CGF}
G(\lambda) =&   \int_0^\mathcal{T} dt\ln(\det(M_{t,0} + \tilde{M}_t))
\\
=& \ G^{\mathrm{elas}}(\lambda) + G^{\mathrm{pump}}(\lambda)
\end{aligned}
\end{equation}
where
\begin{equation}
\begin{aligned}
\nonumber
 G^{\mathrm{elas}}(\lambda)&=\int_0^\mathcal{T} dt \ln(\det(M_{t0})), 
 \\ 
  G^{\mathrm{pump}}(\lambda) &= \int_0^\mathcal{T} dt \Tr \Big( \ln(\mathcal{I}+M_{t,0}^{-1}\tilde{M}_t)\Big),
\end{aligned}
\end{equation}
where $\mathcal{I}$ is the identity matrix. Here we have labeled the contribution which would survive in the static limit and arises due to only elastic scattering events as $G^{\mathrm{elas}}(\lambda)$ and the contribution arising from the adiabatic driving as $G^{\mathrm{pump}}(\lambda)$. Since we are working in the limit of small amplitude parameter modulation, in which these dynamic contributions to the scattering matrix are small, keeping only terms quadratic in $X_{j,\omega}$ would appear to be a justifiable approximation. Terms of this nature appear in both the linear and quadratic contributions to the expansion of the matrix $\ln(M_{t,0}^{-1}\tilde{M}_t)$. Consequently, the contribution to the CGF from the pump can be Taylor expanded and truncated to quadratic order:
\begin{equation}
 G^{\mathrm{pump}}(\lambda) \approx \int_0^\mathcal{T} dt \Tr \Big (M_{t,0}^{-1}\tilde{M}_t - \frac{1}{2}\big (M_{t,0}^{-1}\tilde{M}_t \big )^2 \Big ).
\end{equation}
This approximation has significantly simplified our calculation. In particular the matrix $M_{t,0}$ is block diagonal in the discretised energy basis. Its determinant can then be written as a product of the determinant of each of the individual blocks $M_{t,0}(\epsilon)$. In the continuous limit the static contribution to the CGF is then given by 
\begin{equation}
\label{static gen 1}
 G^{\mathrm{elas}}(\lambda) = \int_0^\mathcal{T} dt \int_{-\infty}^{\infty} d\epsilon \ln\big(\det(M_0(\epsilon))\big).
\end{equation} 
Similarly, since the dynamic contribution can be expressed as a trace, we can again consider the diagonal blocks at each energy separately and in the continuous limit we have that
\begin{equation}
\begin{aligned}
\label{dyn gen 1}
G^{\mathrm{pump}}(\lambda) = \int_0^\mathcal{T} dt \int_{-\infty}^{\infty} d\epsilon \Tr \Bigg[ M_{t,0}(\epsilon)^{-1}\tilde{M}_t(\epsilon)
\\
 - \frac{1}{2}\bigg(M_{t,0}^{-1}(\epsilon)\tilde{M}_t(\epsilon) \bigg )^2 \Bigg].
\end{aligned}
\end{equation} 
Eqs. \ref{static gen 1} and \ref{dyn gen 1} constitute the first main results of this work and can be used to determine the heat transport statistics and fluctuation theorems for weak and adiabatic cyclically driven system in terms of the scattering matrix.

In the case that we have the simultaneous variation of just two parameters of the Hamiltonian, the dynamic contribution to the generating function exhibits two distinct contributions. The first consists of terms dependent only on the variation of a single Hamiltonian parameter and is hence proportional to $X_{j,\omega}^2$. This contribution is independent upon the direction of the driving in parameter space and survives in the case where only a single parameter is varied. The second contribution is, in contrast, geometric in nature and hence only dependent upon the path traversed in parameter space during the driving cycle. We find that this contribution, $G^{\mathrm{geom}}(\lambda)$, is independent of the driving frequency and identified by its proportionality to $X_{1,\omega}X_{2,\omega}$. The sign of this contribution is sensitive to the direction of traversal of the contour in parameter space associated with the driving. Geometric contributions to the full counting statistics of heat transport have previous been demonstrated to produce correction to fluctuation theorems \cite{Ren2010}. This contribution takes on further interest within systems where the accumulated geometric phase is topologically protected against fluctuations in the driving cycle, such as a Majorana braiding protocol. Furthermore, although the derivation of the full counting statistics here used the approximation that the amplitude of the driven cycle is small in parameter space, we show in Sec.\ \ref{Large Amp Drive} that our analysis can be extended to large amplitude pumps for the case of such geometric contributions.

\subsection{Full Counting Statistics for Pumped Charge Transport}
The calculation in the previous section can be reproduced for the case of electronic transport of an adiabatically driven system. In this case, the characteristic function is given by
\begin{equation}
\chi_{e,t}(\lambda) = \Big < e^{i\lambda \hat{Q}_{e,\rightarrow}}e^{-i\lambda \hat{Q}_{e\leftarrow}} \Big >,
\end{equation}
with ingoing and outgoing charge operators defined as $\hat{Q}_{e,\rightarrow(\leftarrow)} = \sum_{\epsilon_i}  e\Big (\hat{N}^e_{\epsilon_i \rightarrow(\leftarrow)}-\hat{N}^h_{\epsilon_i \rightarrow(\leftarrow)} \Big )$ and where $e$ is the unit of electronic charge. This expression reflects the fact that electrons and holes, traveling in the same direction with respect to the scattering centre, carry charge in opposite directions. From this new starting point, one can show that the corresponding elastic and dynamic contributions to the CGF can be expressed analogously to those for the case of heat transport:
\begin{equation}
\begin{aligned}
\label{elec GF}
 G_{e}^{\mathrm{elas}}(\lambda) =& \int_0^\mathcal{T} dt \int_{-\infty}^{\infty} d\epsilon \ln\big(\det(M_{0}^e(\epsilon))\big).
 \\
 G_{e}^{\mathrm{pump}}(\lambda) =& \int_0^\mathcal{T} dt \int_{-\infty}^{\infty} d\epsilon \Tr(M_{t,0}^e(\epsilon)^{-1}\tilde{M}^e_t(\epsilon)),
 \end{aligned}
\end{equation} 
where now
\begin{equation}
\begin{aligned}
& M_{t,0}^e =  1 - \rho
\\
&+\rho \exp(i \lambda \sum_i  P_{\epsilon_i \rightarrow})\Bigg(1 + \sum_i P_{\epsilon_i \leftarrow}^0 (e^{-i \lambda }-1)\Bigg), 
\\
&\tilde{M}_t ^e=   \rho \exp(i \lambda \sum_i  P_{\epsilon_i \rightarrow}) \Bigg ( \sum_i  \tilde{ P}_{\epsilon_i \leftarrow} (e^{-i \lambda }-1) \Bigg ).
\end{aligned}
\end{equation}
In this case, $ P_{\epsilon_i \rightarrow (\leftarrow)} =  P^e_{\epsilon_i \rightarrow (\leftarrow)} -  P^h_{\epsilon_i \rightarrow (\leftarrow)}$ and the matrices $ P^{e(h)}_{\epsilon_i \rightarrow (\leftarrow)}$ are defined as in Eq. \ref{ingoing P} and \ref{outgoing P}. As for the analogous expressions for heat transport, these results are valid for any cyclically driven system, provided the driving can be considered adiabatic and with an amplitude that is small in the relevant parameter space.

%%%%%%%%%%%%%%%%%%%%%%%%%%%%%%%%%%%%%%%%%%%%%%%%%%%%%%%%%%%%%%%%%%%%%%%%%%%%%%%%%%%%%%%%%%%%%%%%%%%%%%%%%%%%%%%%%%%%%%%%%%%%%%%%%%%%%%%%%%%%%%%%%%%%%%%%%%%%%%%%%%%%%%%%%%%%%%
\section{Driven 1D topological superconductors: Braiding Cycle and Pumped Heat}
\label{Braiding Sec}
\begin{figure}
	\centering
	\includegraphics[width=0.45\textwidth]{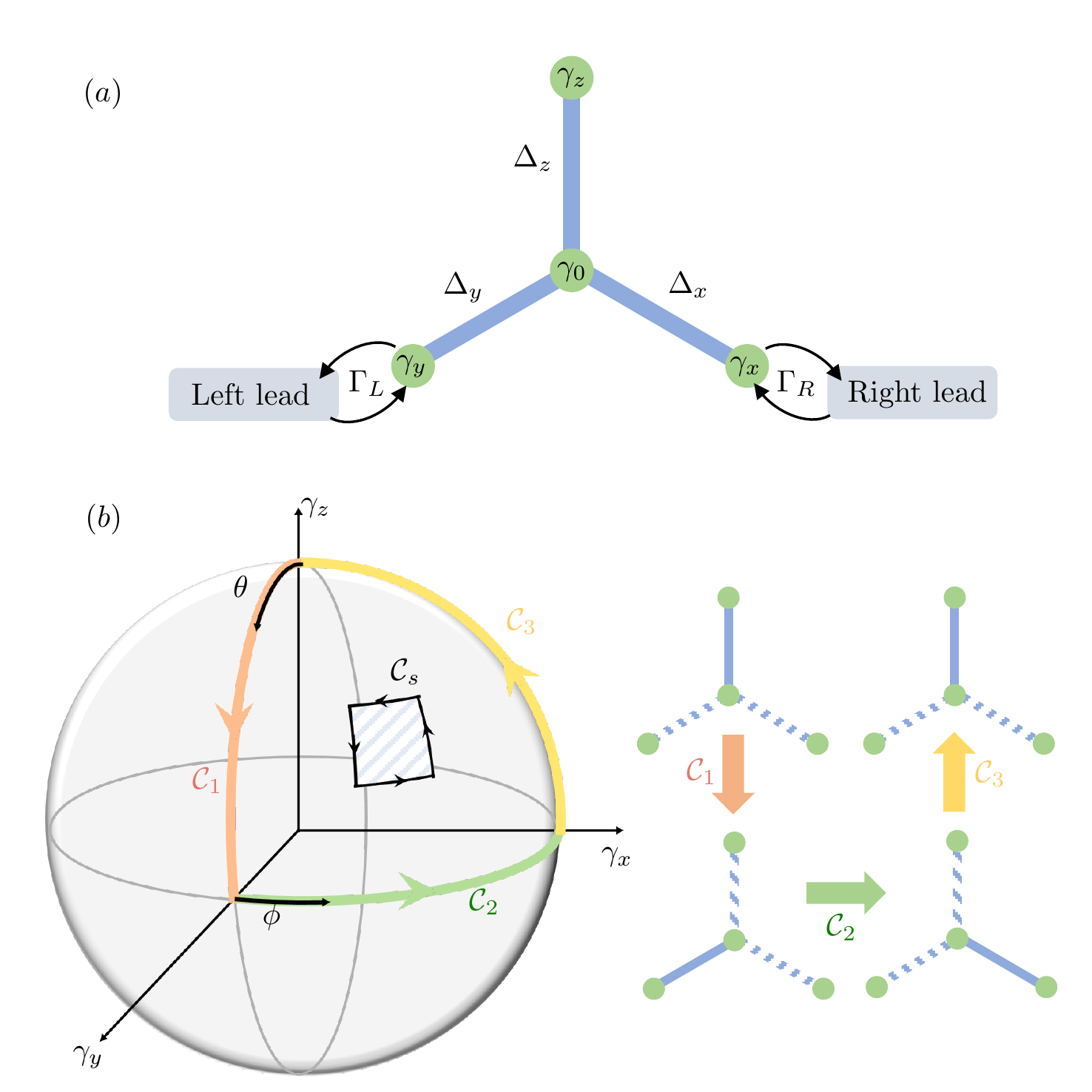} 
	\caption{(a) Y-junction of p-wave superconducting nanowires (blue) with Majorana zero modes at positions indicted by the green dots. Each of the external Majorana modes, $\gamma_{x,y,z}$, is coupled to central mode with corresponding coupling strengths $\Delta_{x,y,z}$ and the modes $\gamma_x$ and $\gamma_y$ are further coupled to conducting normal metal leads with strengths $\Gamma_L$ and $\Gamma_R$. (b) Schematic depicting the Majorana braiding cycle. The diagram on the right illustrates the required sequence of Majorana couplings where the solid blue lines illustrate the couplings which are turned on and dashed lines indicate those that are turned off. The corresponding evolution, $\mathcal{C}_1+\mathcal{C}_2+\mathcal{C}_3$, is shown as a path in spherical parameter space on the left. Also illustrated is an example of a small amplitude driving contour $\mathcal{C}_s$.}
	\label{Setup}
\end{figure}
The driven process, for which we would like to study transport statistics and hence will provide the central focus of this paper, is that of a Majorana braiding. The setup under consideration consists of three p-wave superconducting nanowires, in a topologically nontrivial state, arranged in the form of a Y-junction as illustrated in Fig. \ref{Setup}. Such a system has been demonstrated, by D. Meidan \textit{et al.}, to produce a net heat current under Majorana braiding \cite{Meidan2019}.
Each of the three wires hosts two zero energy Majorana modes, one at each end of the wire \cite{Kitaev2001}. However, at energies well below the superconducting gap, $\Delta_{sc}$, the low energy Hilbert space is spanned by the three outer Majorana zero modes, $\gamma_{x}$, $\gamma_{y}$, $\gamma_{z}$, and a fourth zero mode, $\gamma_{0}$, formed by a linear combination of the internal Majoranas from each wire. This Y-junction is coupled to two external normal metal leads, $L$ and $R$, which lie on either side of the junction in order to facilitate a particle current and allow the exploration of the transport properties of the braiding protocol.

The two Majorana states $\gamma_{x}$ and $\gamma_{y}$ can be exchanged by systematically modulating the couplings between the external Majorana states and the central state $\gamma_{0}$ \cite{VanHeck2012,Karzig2016a}. The couplings between these four states can be quantified in the effective Hamiltonian for the Y-junction,
\begin{equation}
H_{Y} = i\gamma_{0} \vec{\Delta} \cdot \vec{\gamma}, 
\end{equation}
where $\vec{\Delta} = \Delta(\sin\theta \cos\phi,\sin\theta \sin\phi,\cos\theta)$ and $\vec{\gamma} = (\gamma_{x},\gamma_{y},\gamma_{z})$. The complete Hamiltonian for the system is then given by $H=H_{Y}+H_{coup}+H_{leads}$ where the contributions from the coupling to the external leads and the leads themselves can be written as
\begin{equation}
\label{coup H}
\begin{gathered}
H_{\mathrm{coup}} = \sqrt{\Gamma_{L}}(c_{Lk}-c_{Lk}^{\dagger})\gamma_{x} + \sqrt{\Gamma_{R}}(c_{Rk}-c_{Rk}^{\dagger})\gamma_{y},
\\
H_{\mathrm{leads}} = \sum_{k}\sum_{\alpha=L,R} \xi_{\alpha k}c_{\alpha k}^{\dagger} c_{\alpha k},
\end{gathered}
\end{equation}
respectively. Here, $\Gamma_{L/R}$ denote the coefficients associated with particle tunneling from the leads onto the superconducting Y-junction and $\xi_{\alpha k}$ are the energy dispersion relations in the leads. 

The process of braiding now corresponds to adiabatically changing the parameters $\theta$ and $\phi$ in such a way as to generate the evolution of the Majorana couplings, as illustrated in Fig.\ \ref{Setup}(b).

This evolution can be better understood by writing the Hamiltonian, $H_{Y}$, in terms of the basis vectors for a spherical coordinate system.
\begin{equation}
\begin{gathered}
\label{spherical Majoranas}
H_{Y} = i\Delta \gamma_{0} \gamma_{r},
\\ \mathrm{where} \  \gamma_{r} = \vec{\gamma} \cdot \hat{e}_{r}, \ \gamma_{\theta} = \vec{\gamma} \cdot \hat{e}_{\theta}, \ \gamma_{\phi} = \vec{\gamma} \cdot \hat{e}_{\phi}.
\end{gathered}
\end{equation}
Since they do not enter the Hamiltonian, the basis vectors $\hat{e}_\theta$ and $\hat{e}_{\phi}$ span the degenerate ground space and the adiabatic evolution of the system can now be interpreted as changing the projection of the physical Majorana states on to this degenerate ground space. At energies well below the superconducting gap ($\epsilon \ll \Delta_{sc}$), it is only this subspace that will facilitate particle transport between the left and right leads, via the occupation of the zero energy, non-local Fermi level defined via the annihilation operator,
\begin{equation}
\hat{a} = \frac{1}{2} \big( \gamma_\theta + i \gamma_\phi \big).
\end{equation}
It has been demonstrated \cite{Karzig2016a,VanHeck2012} that the sequence of couplings sketched in Fig.\ \ref{Setup}(b) corresponds to this annihilation operator accumulating a phase factor $e^{i \Omega_\mathcal{C}}$, where $\Omega_\mathcal{C}$ corresponds to the solid angle enclosed by the curve $\mathcal{C}=\mathcal{C}_1+\mathcal{C}_2+\mathcal{C}_3$ traversed in parameter space. For the specific process outlined in this section with, $\Omega_\mathcal{C} = \pi/2$, the resulting unitary evolution operator, $U = e^{-\frac{\pi}{4}\gamma_\phi \gamma_\theta}$, corresponds to the exchange of Majoranas $\gamma_x$ and $\gamma_y$:
\begin{eqnarray}
\nonumber
U^\dagger \gamma_x U = \gamma_y,
\\
\nonumber
U^\dagger \gamma_y U = -\gamma_x.
\end{eqnarray}     

This braiding protocol also leads to a non-zero heat current being pumped between the two external leads and, in the low temperature limit, the total heat pumped tends to some universal value independent of the coupling strength to the external leads and fluctuations to the driving \cite{Meidan2019}. Despite this, the particle-hole symmetry of the Majorana-lead coupling results in no transfer of charge between the leads during the process. 

In order to find the CGF for the Majorana braiding protocol, we first need to determine its instantaneous scattering matrix $S(\epsilon,t)$. For the superconducting Y-junction, the scattering matrix can be calculated as \cite{Mahaux1969}:
\begin{equation} 
\label{scatter def}
S(\epsilon,t) = 1 + 2\pi i W^{\dagger} (H_0-\epsilon-i\pi WW^\dagger )^{-1}W.
\end{equation}
Where $H_{0}$ denotes the Hamiltonian of the Y-junction in the absence of the external leads and $W$ describes the coupling between the incoming electron and hole (e/h) scattering states in the leads and the states of the system. This coupling matrix can be obtained from the Hamiltonian in Eq. \ref{coup H} and, in the basis of the Majorana zero modes, it takes the form:
\begin{equation}
\begin{aligned}
W = \sqrt{\Gamma_{L}} \Big (\ket{\gamma_x}\bra{e^L}-&\ket{\gamma_x}\bra{h^L}\Big ) 
\\
&+ \sqrt{\Gamma_{R}} \Big (\ket{\gamma_y}\bra{e^R}-\ket{\gamma_y}\bra{h^R}\Big ).
\label{eq:coupling matrix} 
\end{aligned}
\end{equation}
Eq. \ref{scatter def} and \ref{eq:coupling matrix} then give the specific form of the scattering matrix for the braiding protocol \cite{Meidan2019}:
\begin{equation}
\label{specific matrix 1}
\begin{gathered}
S(\epsilon) = \begin{pmatrix}
S^{L^eL^e} & 1-S^{L^eL^e} & S^{L^eR^e} & -S^{L^eR^e} \\
1-S^{L^eL^e} & S^{L^eL^e} & -S^{L^eR^e} & S^{L^eR^e} \\
S^{L^eR^e} & -S^{L^eR^e} & S^{R^eR^e} & 1-S^{R^eR^e} \\
-S^{L^eR^e} & S^{L^eR^e} & 1-S^{R^eR^e} & S^{R^eR^e} \\
\end{pmatrix},
\\
\mathrm{where} \ \ S^{L^eL^e} = 1 - 4\pi it
\Gamma \Bigg (\frac{\sin^2\phi}{\epsilon + 2\pi i\Gamma} + \frac{\cos^2\theta \cos^2\phi}{\epsilon + 2\pi i\cos^2\theta \Gamma} \Bigg ),
\\
S^{R^eR^e} = 1 - 4\pi i\Gamma \Bigg (\frac{\cos^2\phi}{\epsilon + 2\pi i\Gamma} + \frac{\cos^2\theta \sin^2\phi}{\epsilon + 2\pi i\cos^2\theta \Gamma} \Bigg ),
\\
\mathrm{and} \ \ S^{L^eR^e} = \frac{4\pi i\epsilon\cos\phi\sin^2\theta\sin\phi \Gamma_R^2}{(\epsilon+2\pi i\Gamma)(\epsilon+2\pi i\cos^2\theta \Gamma)},
\end{gathered}
\end{equation}
where here we display the case of equal coupling to the left and right leads, $\Gamma_L=\Gamma_R=\Gamma$, for brevity. The components of this scattering matrix can be used in Eq. \ref{static gen 1} and \ref{dyn gen 1} in order to determine the heat and charge transfer statistics of a driven Majorana Y-junction.

\begin{figure}
	\includegraphics[width=0.45\textwidth]{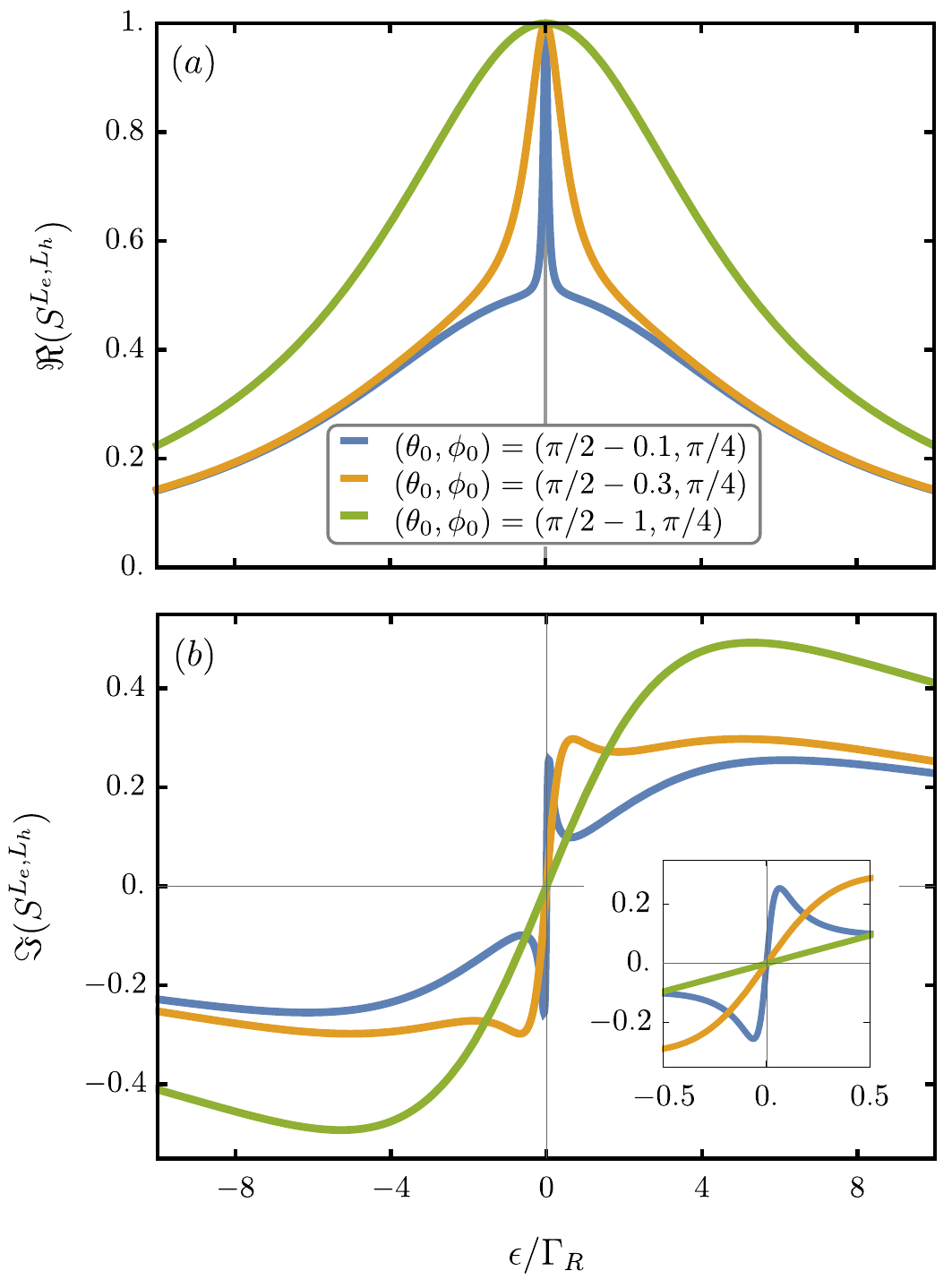}
	\caption{(a) Real and (b) imaginary parts of the Andreev reflection component of the scattering matrix for the topological superconducting Y-junction. Results are plotted for several positions in the parameter space, $(\theta_0,\phi_0)$, and for equal coupling to the left and right external leads $\Gamma_L=\Gamma_R=1$.}
	\label{fig:Scat}
\end{figure}

From the form of the scattering matrix components it is evident that there exists two distinct energy scales that will influence the transport. These scales are illustrated in Fig.\ \ref{fig:Scat} for the case of the Andreev reflection component of the scattering matrix, $S^{L_e,L_h}(\epsilon)=1-S^{L_e,L_e}(\epsilon)$. In Fig.\ \ref{fig:Scat}(a) we see that the energy dependence of the Andreev reflection via the Majorana zero modes consists of the sum of two peaks centred at $\epsilon=0$. The width of the narrower peak, $\Gamma_R \cos^2\theta $, is set by the position in the parameter space $(\theta,\phi)$, with the width decreasing as we approach the line $\theta=\pi/2$, corresponding to the equator of the spherical parameter space shown in Fig.\ \ref{Setup}(b). This energy scale is not visible as we move a sufficient distance away from this line and the larger energy scale dominates. The size of this larger energy scale is set by the strength of the coupling to the external leads $\Gamma_{L/R}$. 

It is also worth noting the difference in behaviour between the real and imaginary components of the scattering matrix in the limit $\epsilon \to 0$. Whereas, the real part can be approximated as constant in this limit, the imaginary part varies linearly with energy and hence quantities that include this contribution will show sensitivity to the energy dependence of the scattering matrix even in the limit $T \to 0$. 
%%%%%%%%%%%%%%%%%%%%%%%%%%%%%%%%%%%%%%%%%%%%%%%%%%%%%%%%%%%%%%%%%%%%%%%%%%%%%%%%%%%%%%%%%%%%%%%%%%%%%%%%%%%%%%%%%%%%%%%%%%%%%%%%%%%%%%%%%%%%%%%%%%%%%%%%%%%%%%%%%%%%%%%%%%%%%%
\section{Heat and charge transport cumulants in small cycles}
\label{sec:transport}
\begin{figure*}
	\includegraphics[width=0.9\textwidth]{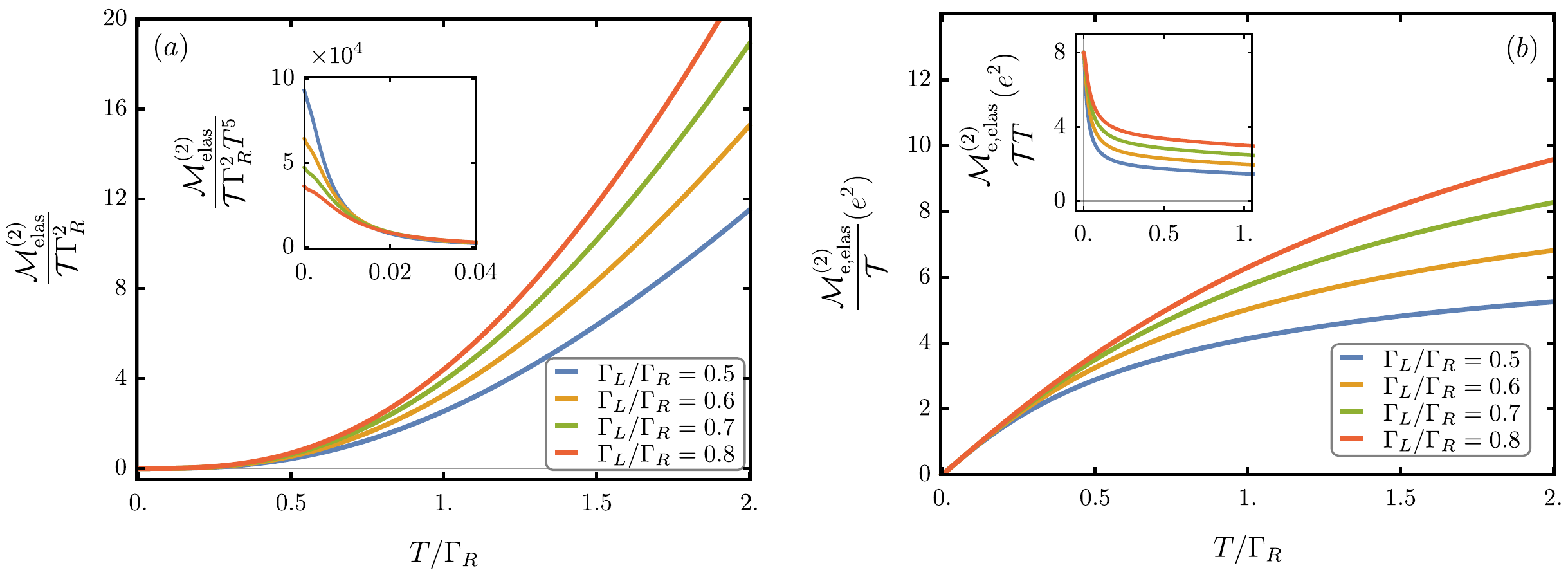}
	\caption{Period-averaged static contribution to the second cumulant of (a) the pumped heat and (b) pumped charge throughout the driving of a Majorana Y-junction centred at $(\theta_0,\phi_0) = (\pi/2-0.1,\pi/4)$, with amplitude $\theta_\omega = \phi_\omega = 0.01$. The noise is plotted as a function of the external lead temperature $T/\Gamma_R$, for a driving frequency $\omega/\Gamma_R=0.001$. The insets show the temperature dependence of this quantity scaled by $T^5$ and $T$ for heat and charge respectively, highlighting the behaviour as $T \to 0$. The different colours correspond to various values of the coupling between the Y-junction and the external leads $\Gamma_L/\Gamma_R$ (cf. legend).}
	\label{fig:elas heat noise}
\end{figure*}
With the motivation of studying the FCS for a Majorana braiding protocol, we begin by considering a situation where the superconducting Y-junction, outlined in Sec. \ref{Braiding Sec}, is driven through a small amplitude cycle on the surface of the spherical phase space defined by the parameters $\theta$ and $\phi$ and illustrated by the contour $\mathcal{C}_s$ in Fig. \ref{Setup}(b).
The CGF for this process is found by substituting the scattering matrix, defined in Eq.\ \ref{specific matrix 1}, into the elastic and dynamic contributions (Eq. \ref{static gen 1} and \ref{dyn gen 1}) at some time $t$. For simplicity, we will consider the case in which we have no chemical potential bias, $\mu_L=\mu_R=0$, between the external leads and that both leads are held at the same temperature, $T_L=T_R=T$, so that the distribution functions for holes and electrons in each lead are identical: $f_\mathrm{in}^{e,L}(\epsilon) = f_\mathrm{in}^{h,L}(\epsilon) = f_\mathrm{in}^{e,R}(\epsilon) = f_\mathrm{in}^{h,R}(\epsilon)$. Our approach is valid in the adiabatic limit, which in this case corresponds to restricting the driving frequency to be much smaller than the coupling between the system and the external leads, $\omega \ll \Gamma_{L,R}$.

The expressions for the generating function in Eq. \ref{static gen 1} and \ref{dyn gen 1} can be used to determine all cumulants of both the heat, $\mathcal{M}^{(k)}$, and charge, $\mathcal{M}^{(k)}_{e}$ transport between the external leads:
\begin{equation}
\mathcal{M}^{(k)}_{(e)} = \frac{\partial^k G_{Q_{(e)}}(\lambda)}{\partial (i \lambda)^k}\Bigg \rvert _{\lambda=0}.
\end{equation}
For example the first cumulants $\mathcal{M}^{(1)}$ and $\mathcal{M}^{(1)}_{e}$ correspond to the total heat, $\big<\hat{Q}(t)\big>$ and charge, $\big<\hat{Q}_{e}(t)\big>$  pumped during the cycle respectively, and the second cumulants $\mathcal{M}^{(2)}_{(e)}$ give the variances, or noise, $ \big<\hat{Q}^2_{(e)}(t)\big> - \big<\hat{Q}_{(e)}(t)\big>^2$ of these distributions.

% \mynotes{the noise defined as the zero Fourier coefficient of the two time current correlation function:
%\begin{equation}
%\mathcal{M}^{(2)}_{(e)} = \int \frac{dt}{\mathcal{T}} %\int_0^\mathcal{T} dt' \mathcal{S}_{(e)}(t,t'),
%\end{equation}
%where $\mathcal{S}_{(e)}(t,t') = \Big < \Delta\hat{I}_{(e)}(t) %\Delta\hat{I}_{(e)}(t') \Big >$
%and $ \Delta\hat{I}_{(e)} = \hat{I} - \big<\hat{I}_{(e)}\big> $. }

\subsection{Elastic contribution to energy and charge transfer statistics}
We first discuss the contribution to the CGF arising from elastic scattering events only, which would survive in the limit that the system is not driven and is hence relevant for cycles of arbitrary amplitude. For the case of heat transport we find that the static contribution can be expressed as
\begin{equation}
\label{static G 1}
\begin{aligned}
G^{\mathrm{elas}}(\lambda) =   \mathcal{T} \int_{-\infty}^{\infty} d\epsilon \ln(1 + \sum_{n=-1}^{1} B_n(\epsilon) (e^{i\lambda \epsilon n}-1)),
\end{aligned}
\end{equation} 
where $\mathcal{T}$ denotes the period of the driving and the coefficients $B_n(\epsilon)$ take the form
\begin{equation}
\begin{gathered}
B_{1}(\epsilon)=B_{-1}(\epsilon) = 4|S^{L^e,R^e}(\epsilon,\theta_0,\phi_0)|^2 f(\epsilon) \big ( 1-f(\epsilon) \big ).
\end{gathered}
\end{equation}
Here, $(\theta_0,\phi_0)$ corresponds to the location of the driving path centre in parameter space. In this form the physical meaning of the CGF becomes clear, as heat is only transferred across the junction by the normal and Andreev transmission of electrons and holes in both directions. For example, the transmission of an electron from the left to the right lead will occur with a probability of $|S^{L^eR^e}(\epsilon)|^2 f(\epsilon)(1-f(\epsilon))$, as expected. The corresponding expression for charge transport is found to be
\begin{equation}
\label{static G Heat}
G_{e}^{\mathrm{elas}}(\lambda) =   \mathcal{T}\int_{-\infty}^{\infty} d\epsilon \ln(1 + \sum_{n=-2}^{2} B_n(\epsilon) (e^{i\lambda n}-1)),
\end{equation}
where 
\begin{equation}
\nonumber
\begin{gathered}
\ \ B_{-1}(\epsilon) = B_1(\epsilon) = 4|S^{L^eR^e}(\epsilon,\theta_0,\phi_0)|^2 f(\epsilon) \big ( 1-f(\epsilon) \big ),
\\
\ \ B_{-2}(\epsilon) = B_{2}(\epsilon) = |S^{L^eL^h}(\epsilon,\theta_0,\phi_0)|^2f(\epsilon) \big ( 1-f(\epsilon) \big ).
\end{gathered}
\end{equation}
Here we see the additional contribution of Andreev reflection processes, which  result in the creation and annihilation of Cooper pairs within the superconducting nanowire system. These processes result in the propagation of an electronic charge of $\pm 2e$ but no energy transport in the form of heat. 

In the case of both zero temperature and chemical potential bias between the external leads, both the heat and charge current contributions arising from the elastic CGF, $ G_{(e)}^{\mathrm{elas}}(\lambda)$, are identically zero,  $\big<\hat{Q}_{\mathrm{elas}} \big> = \big<\hat{Q}_{e,\mathrm{elas}}\big>=0$. Despite this, the elastic scattering processes still allow for fluctuations. This contribution to the noise is thermal in nature, arising due to thermal fluctuations in the occupation of the ingoing scattering states. It vanishes in the limit $T=0$ and $\mu=0$  when the occupation of all ingoing energy states is fixed and no charge or energy transfer processes take place. This thermal noise is present in both energy and charge transport and for our setup it is natural to identify two distinct temperature regimes relative to the strength of coupling to the external metal leads.
%%%%%%%%%%%%%%%%%%%%%%%%%%%%%%%%%%%%%%%%%%%%%%%%%%%%%%%%%%%%%%%%%%%%%%%%%%%%%%%%%%%%%%%%%%%%%%%%%%%%%%%%%%%%%%%%%%%%%%%%%%%%%%%%%%%%%%%%%%%%%%%%%%%%%%%%%%%%%%%%%%%%%%%%%%%%%%
\subsubsection{Thermal noise at low temperature} 
For low values of the external lead temperature, $T\ll \min \Gamma_{L/R}$, the energy dependence of the absolute value of the scattering matrix elements can be considered weak (cf.\ Fig.\ \ref{fig:Scat}). Consequently, one would expect that the behaviour of the thermal noise as a function of temperature in this regime should be dictated by the Fermi distribution functions, $f(\epsilon)$, appearing in the elastic contribution to the CGF (Eq.\ \ref{static G 1}, \ref{static G Heat}). Taking the scattering matrix elements to be energy independent, the form of the static contribution to the CGF implies that the elastic charge noise should depend linearly on temperature, whereas the elastic heat noise should vary as $T^3$. This behaviour is well understood and in agreement with previous studies of transport statistics. \cite{Moskalets2002,Moskalets2002a}      

The period averaged second cumulants of the static contribution to the CGF, $\frac{1}{\mathcal{T}}\mathcal{M}_{(e),\mathrm{elas}}^{(2)}$, which quantify the thermal noise, are plotted for both heat and charge in Fig. \ref{fig:elas heat noise}(a) and \ref{fig:elas heat noise}(b) respectively. At low temperatures $T\ll \min \Gamma_{L/R}$, we see that, as expected, the electronic charge thermal noise scales linearly with temperature. Additionally, we see that the charge noise becomes independent of the coupling to the external leads, $\Gamma_{L/R}$ (cf.\ Fig.\ \ref{fig:elas heat noise}(b) inset). This is a further consequence of the weak energy dependence of the frozen in time scattering matrix, $\hat{S}(\epsilon,X_{j,0})$ at energies close to zero.

We find that the thermal heat noise, however, is sensitive to the energy dependence of the scattering matrix even in the low temperature limit. In fact, if the energy dependence was neglected entirely, and the scattering matrix evaluated at the chemical potential $\mu=0$, the elastic contribution to the heat noise would vanish. The inset in Fig.\ \ref{fig:elas heat noise}(a) illustrates that as $T \to 0$ the elastic heat noise scales as $T^5$ as opposed to the $T^3$ scaling originating from the distribution functions of the normal metal leads. The influence of the scattering matrix is also evident in the fact that the thermal noise is dependent upon the external lead coupling at all temperatures, in contrast to the case of charge transport. However, the energy dependence of the scattering matrix cannot be entirely neglected even in the case of charge transport, as its influence is evident in the contributions to the transport cumulants arising from the pumping.

\subsubsection{Thermal noise at high temperature}
\begin{figure*}
	\includegraphics[width=1\textwidth]{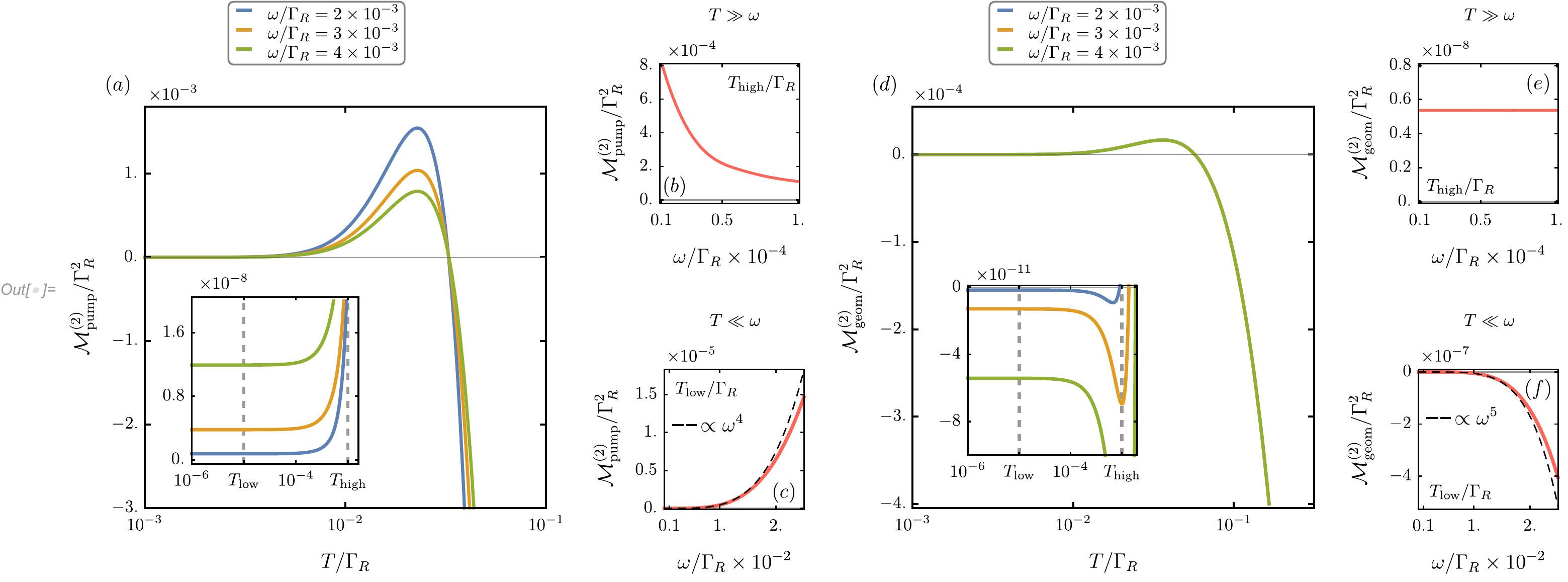}
	\caption{The pumped contribution to the second cumulant of the heat transport throughout the driving of a Majorana Y-junction centred at $(\theta_0,\phi_0)=(\pi/2-0.1,\pi/4)$ with amplitude $\theta_\omega=\phi_\omega=0.01$. Plots $(d,e,f)$ show the geometric contribution whereas $(a,b,c)$ illustrate the remaining non-geometric part. Plots $(a,d)$ show the second cumulants as a function of temperature, with the inset highlighting the region $T \ll \omega$. Panels $(b,c,e,f)$ show the same quantities plotted against frequency. $(b,e)$ illustrate the behaviour as a function of low frequencies $\omega < T$ and $(c,f)$ at high frequencies $\omega > T$.}
	\label{fig: pumped heat noise}
\end{figure*}
As the temperature becomes comparable with $\min \Gamma_{L/R}$, the energy dependence of the scattering matrix becomes increasingly significant in both the cases of charge and heat noise. For scattering processes via Majorana zero modes, transport is dominated by low energy scattering states, hence high energy occupation fluctuations that occur as $T$ is increased do not contribute to the current noise. As a result, the rate at which the elastic noise increases slows down at high temperatures and will eventually plateau for the case of charge transport and scale $\propto T$ for that of heat transport. The temperature at which this occurs is proportional to the external lead coupling $\min \Gamma_{L/R}$, as can be clearly seen in the main panels of Figs.\ \ref{fig:elas heat noise}a and \ref{fig:elas heat noise}b for heat and charge noise respectively.

It is also worth noting that, at all temperatures both the period averaged charge and heat thermal noises are independent of the driving frequency. This is to be expected, as the components of the scattering matrix responsible for elastic scattering events are not influenced by the driving.
%%%%%%%%%%%%%%%%%%%%%%%%%%%%%%%%%%%%%%%%%%%%%%%%%%%%%%%%%%%%%%%%%%%%%%%%%%%%%%%%%%%%%%%%%%%%%%%%%%%%%%%%%%%%%%%%%%%%%%%%%%%%%%%%%%%%%%%%%%%%%%%%%%%%%%%%%%%%%%%%%%%%%%%%%%%%%%
\subsection{Averaged pumped heat and energy}
Next, we analyse the more interesting contributions to the transport statistics arising from the pumped contribution to the CGF given in Eq. \ref{dyn gen 1}. Again considering the case for which the external leads are held at the same temperature, $T$, and zero chemical potential $\mu=0$, one finds that the charge pumped during any modulation of the Majorana Y-junction is identically zero. This is a direct result of the electron-hole symmetry of the coupling between the Majorana zero modes and the leads \cite{Meidan2019}. This result is in contrast to previous works on adiabatic pumps in which the scattering matrix does not possess such a symmetry and the pumped charge is found to vary linearly with the pumping frequency \cite{Moskalets2002}. Despite this there is a finite heat current pumped across the junction which, in the case of zero temperature bias, arises solely from the geometric part of the CGF, $G_Q^{\mathrm{geom}}(\lambda)$:
\begin{equation}
\label{pumped heat}
\begin{aligned}
&\big<Q^{\mathrm{pump}}\big> = \big<Q^{\mathrm{geom}}\big> = \frac{\partial}{\partial (i \lambda)} G^{\mathrm{geom}}(\lambda)\Bigg \rvert _{\lambda=0}
\\
=&\sum_{\beta=eL,eR} 4  \int_{-\infty}^{\infty} d\epsilon \ \epsilon \iint \frac{\partial f(\epsilon)}{\partial \epsilon} \Im \Bigg[\frac{\partial S^{eL\beta}(\epsilon)}{\partial \theta}\frac{\partial S^{eL\beta}(\epsilon)}{\partial \phi}\Bigg] d\theta d\phi.
\end{aligned}
\end{equation}
In this form the geometric nature of the pumped heat becomes clear, since its value depends only upon the contour traversed in parameter space throughout the driving and is independent of the driving frequency itself. The fact that the pumped heat arises solely from the geometric contribution to the CGF means that this expression is valid for arbitrary amplitude cycles in parameter space, (see Sec. \ref{Large Amp Drive}), and in particular for that of the Majorana braiding protocol illustrated in Fig. \ref{Setup}(b). The energy pumped throughout such a process is found to be in agreement with Ref. \cite{Meidan2019}. 

The Majorana braiding is of particular interest since the path traversed in parameter space during the process is topologically protected against fluctuations in the driving. As a consequence, any transport properties that can be shown to be geometric in nature will also be protected. Furthermore, at low temperatures, $T \rightarrow 0$, the derivative of the Fermi function ensures that only particles with energies close to the chemical potential take place in the transport, which in this case corresponds to taking the limit $\epsilon \rightarrow 0$ of the area integral in Eq. \ref{pumped heat}. In this limit the contour traversed in parameter space maps on to a fixed path in scattering matrix space and hence the pumped heat tends to some universal value, independent of the coupling to the external leads as well as the nature of the driving \cite{Meidan2019}:
\begin{equation}
\frac{Q}{2T \log 2} = \frac{1}{4}.
\end{equation} 
%%%%%%%%%%%%%%%%%%%%%%%%%%%%%%%%%%%%%%%%%%%%%%%%%%%%%%
\subsection{Heat and charge noise from pumping}
We now extend our analysis beyond the known results for the average current by analysing the behaviour of higher order cumulants of the contribution to the CGF arising from the time dependent pumping, for transport via Majorana modes. When dynamic process are included, noise can originate not only from thermal fluctuations, but also from the action of the pump itself. This noise arises due to the non-equilibrium nature of the outgoing scattering states, as a consequence of the possibility of scattering events between nearest energy sidebands, and is present in both the cases of heat and electronic transport. This side band scattering results in correlations between outgoing particle distributions at energies within the range $\epsilon \pm \omega$ which manifest themselves as a source of noise in the average pumped heat and charge. This source of noise vanishes in the case that the pump in switched off and hence inelastic scattering events between side bands cease to occur.

From the total pumped noise we can isolate the contribution that arises from the geometric part of the CGF, which we denote $\mathcal{M}^{(2)}_{(e),\mathrm{geom}}$. This is the additional noise that is observed in the case that two parameters are driven simultaneously in a closed cycle. The remaining pumped noise, denoted $\mathcal{M}^{(2)}_{(e),\mathrm{pump}}$,  would be present even in the case that just a single parameter is driven. These pumped contributions to the second cumulant of the heat and charge noise are shown in Fig.\ \ref{fig: pumped heat noise} and \ref{fig: pumped charge noise} respectively. They illustrate the existence of three distinct temperature regimes that dictate the behaviour of the noise as a function of driving frequency $\omega$, each of which are discussed in the following sections.

\subsubsection{Low temperature regime: $T \ll \omega$}
\begin{figure*}
	\includegraphics[width=1\textwidth]{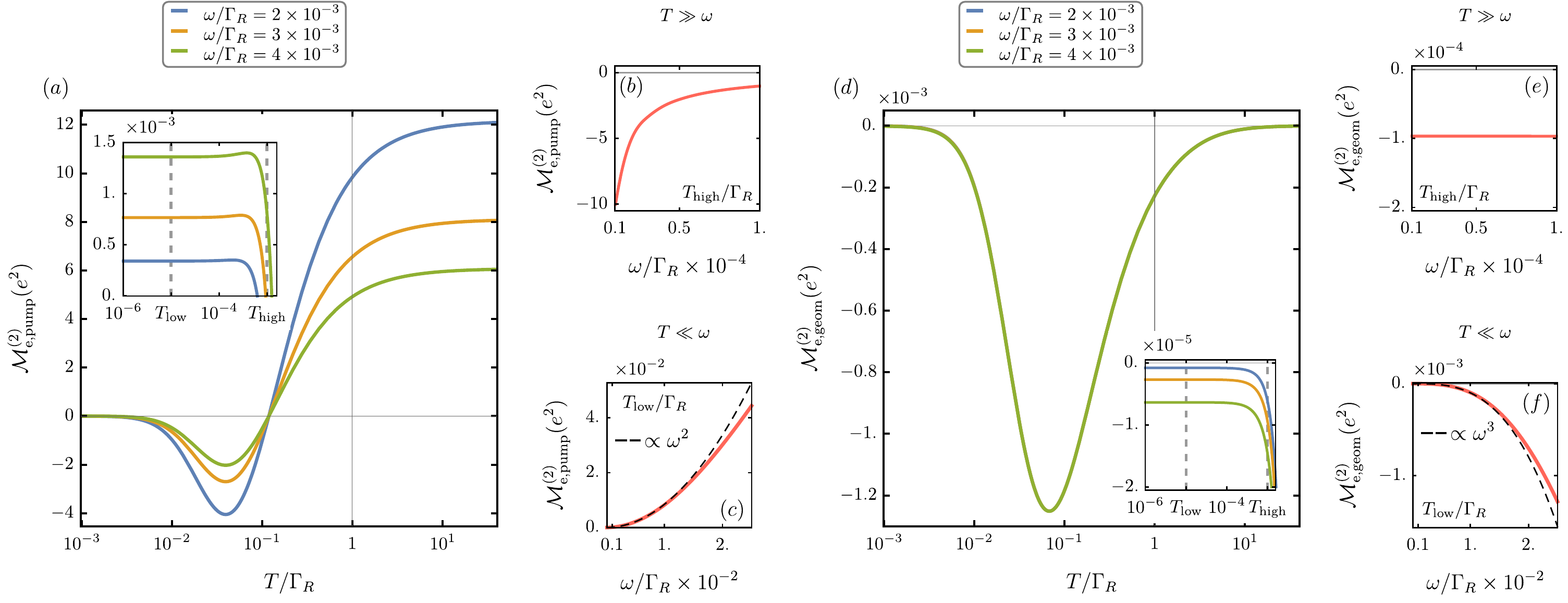}
	\caption{The pumped contribution to the second cumulant of the charge transport throughout the driving of a Majorana Y-junction centred at $(\theta_0,\phi_0)=(\pi/2-0.1,\pi/4)$ with amplitude $\theta_\omega=\phi_\omega=0.01$. Plots $(d,e,f)$ show the geometric contribution whereas $(a,b,c)$ illustrate the remaining non-geometric part. Plots $(a,d)$ show the second cumulants as a function of temperature, with the inset highlighting the region $T \ll \omega$. Panels $(b,c,e,f)$ show the same quantities plotted against frequency. $(b,e)$ illustrate the behaviour as a function of low frequencies $\omega < T$ and $(c,f)$ at high frequencies $\omega > T$.}
	\label{fig: pumped charge noise}
\end{figure*} 
In the case that the temperature is lowered below the energy associated with the driving frequency $\omega$, the thermal noise becomes negligible and the noise associated with the pumping itself dominates. In the case of driven noise, the key quantity dictating the characteristic behaviour is the difference in Fermi occupation functions between neighbouring energy sidebands, of which the pump can stimulate transitions between. In this regime, the quantity $f_0(\epsilon)-f_0(\epsilon \pm \omega)$ is only non-zero over an energy window close to $\epsilon=0$, the width of which scales linearly with $\omega$, but is insensitive to the temperature of the external leads. 

The absence of temperature dependence is reflected in Figs.\ \ref{fig: pumped heat noise}(a,d) and \ref{fig: pumped charge noise}(a,d) for heat and charge transport respectively. The inset panels within each of these figures show that the pumped contributions to the noise tend to some non-zero value in the limit $T \to 0$, in contrast to the static contributions to the noise (cf. Fig. \ref{fig:elas heat noise}). 

Although the energy dependence of the real contribution to the scattering matrix in negligible in this low temperature limit, (cf.\ Fig.\ \ref{fig:Scat}), the linear energy dependence of the imaginary contribution will influence the transport properties, as will the behaviour of the scattering matrix derivatives appearing in the inelastic terms $S_{\pm \omega}(\epsilon)$. This energy dependence manifests itself in the form of a difference in the frequency dependence between the geometric and non-geometric contributions to the pumped noise, since they depend differently on the scattering matrices, as implicit from Eq.\ \ref{GF Full Form}. Specifically, $\mathcal{M}^{(2)}_{\mathrm{pump}}\propto\omega^4$ and $\mathcal{M}^{(2)}_{\mathrm{geom}}\propto\omega^5$ (cf. Fig.\ \ref{fig: pumped heat noise}(c,f)) whereas for charge the non geometric and geometric contributions are found to vary as $\omega^2$ and $\omega^3$ respectively (cf.\ Fig.\ \ref{fig: pumped charge noise}(c,f)). This difference between heat and charge noise can be justified by the fact that the dominant process is the scattering from states of energy $\epsilon$ to $\epsilon\pm \omega$. Consequently, the heat noise is underpinned by the same fluctuations as those in the case of charge, and differ only by the fact that the scattering events are weighted by the energy absorbed/emitted $\sim \omega$.

\subsubsection{Mid-temperature regime: $\omega<T\ll\Gamma_{L,R}$}
As the temperature is increased beyond the driving frequency, the temperature becomes the quantity which determines the energy window within which scattering events can occur. This transition can be seen in panels (a) and (d) of Figs.\ \ref{fig: pumped heat noise} and \ref{fig: pumped charge noise} by the deviation of the pumped noises away from their corresponding constant low temperature values at approximately $T=\omega$. Beyond this point the temperature dependence is governed by a combination of the distribution functions and the energy dependence of the scattering matrix. Initially, the heat noise varies as $T^5$ and we find that the charge noise is proportional to $T^3$, with the ratio $\mathcal{M}^{(2)}/\mathcal{M}^{(2)}_e = T^2$, where the temperature now plays the same role of the frequency in the previous case, in agreement with previous works \cite{Moskalets2004}. However, as $T$ increases further we see that, for both heat and charge, the second cumulant is not monotonic and exhibits a turning point corresponding to the temperature exceeding the width of the scattering matrix resonance. The width of this resonance is set by the location of the centre of the driving in the parameter space, $(\theta_0,\phi_0)$ (cf.\ Fig.\ \ref{fig:Scat}). We also see that the pumped contributions to both the heat and charge noise undergo a sign change which originates from the energy dependence of the derivatives of the scattering matrix with respect to the driving parameters, $S_{\pm \omega}(\epsilon)$, which dictate transitions between nearest energy sidebands. The sum of the static and pumped contributions to the noise, however, remains positive at all temperatures.  
 
The driving frequency dependence of the pumped noise in this regime is similar for both the cases of heat and charge. The difference between the non-geometric and geometric contributions persists however as illustrated in panels (b) and (e) respectively. We see that the non-geometric part is now inversely proportional to $\omega$ whereas the geometric contribution is independent of the frequency of the driving and determined purely by the path traversed in parameter space. 

\subsubsection{High temperature regime: $T>\Gamma_{L,R}$}   
As the temperature is increased beyond the broadening of the scattering matrix resonance, set by the strength of the coupling between the system and the external leads, the scattering matrix dependence on energy is dominated by the generic $1/\epsilon^2$ behaviour. This leads to saturation of charge noise and heat noise that is linear in $T$, (cf.\ panels (a,d) of Figs.\ \ref{fig: pumped heat noise} and \ref{fig: pumped charge noise}).
%%%%%%%%%%%%%%%%%%%%%%%%%%%%%%%%%%%%%%%%%%%%%%%%%%%%%%

%%%%%%%%%%%%%%%%%%%%%%%%%%%%%%%%%%%%%%%%%%%%%%%%%%%%%
\section{Impact upon Fluctuation Theorems}
\label{sec:theorem}
\begin{figure}
	%	\subfigure[]{\includegraphics[width=0.45\textwidth]{"3D GF".pdf}}
	\includegraphics[width=0.45\textwidth]{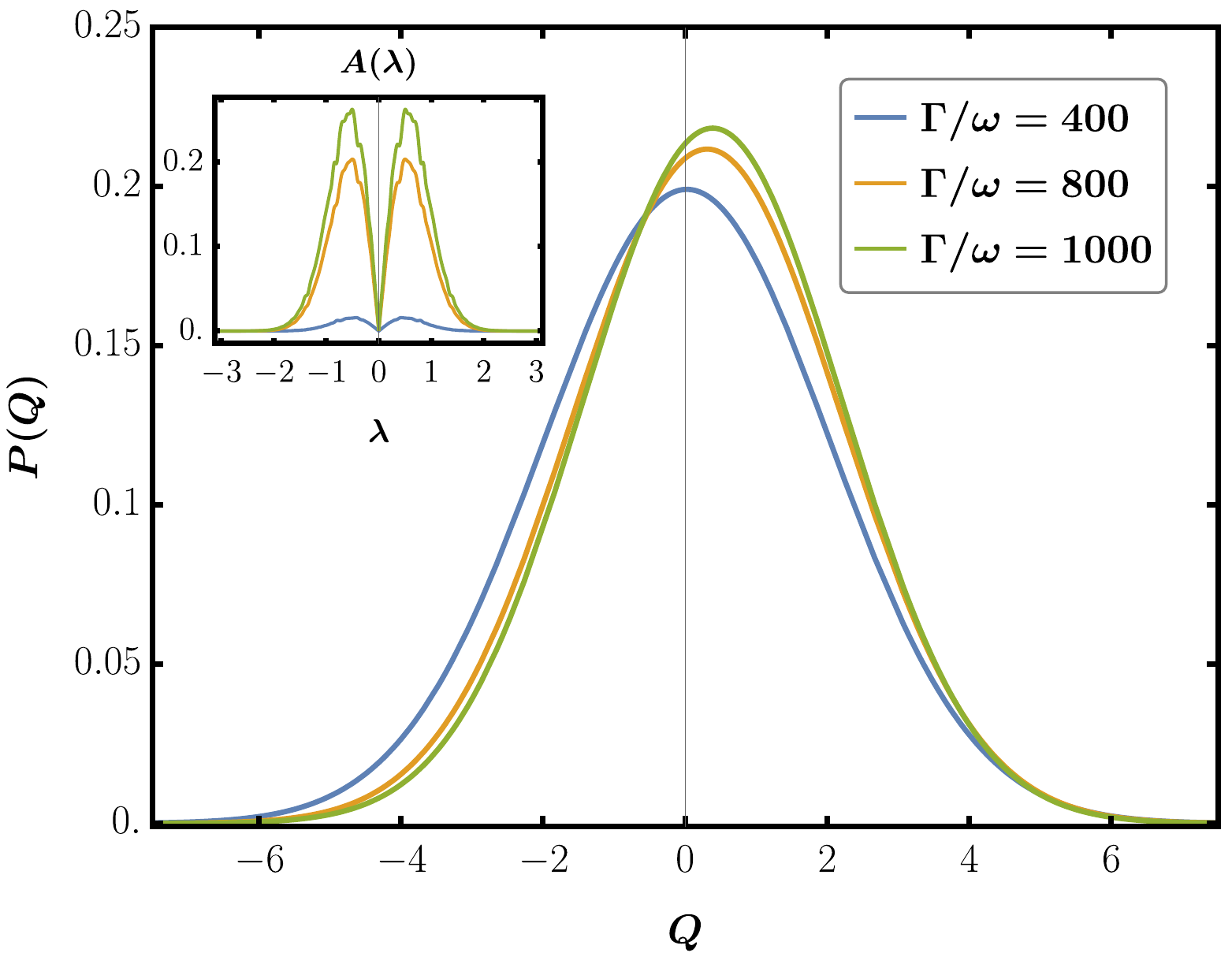}
	\caption{Probability distribution, $P(Q)$, for the heat pumped via the small amplitude ($\theta_\omega=\phi_\omega=0.01$) driving of a Majorana Y-junction centred at $(\theta_0,\phi_0)=(\frac{\pi}{2}-0.01,\frac{\pi}{4})$. Results are shown for several values of the coupling to the external leads, $\Gamma_L=\Gamma_R=\Gamma$, with an external lead temperature of $T/\omega = 10$. The inset shows the corresponding behaviour of the fluctuation theorem violation quantifier $A(\lambda)=|\chi(\lambda)-\chi(-\lambda)|$ which is identically zero when the Gallavotti-Cohen fluctuation theorem holds true.}
	\label{fig:Prob Small Cycle}
\end{figure}
When studying systems which involve heat transfer to thermal reservoirs, fluctuation theorems (FT)  dictate the likelihood of anomalous transfer events which appear in violation with the second law of thermodynamics. Therefore,  they provide useful information on the nature of the heat flow. From their general formulation in terms of entropy production \cite{seifert2005entropy,seifert2012stochastic}, fluctuation theorems can be recast in more specific forms for different settings.  One such example is the Gallavotti-Cohen (GC) FT, determining the statistics of heat transfer between reservoirs at different temperatures. It states that, over a sufficiently long time interval $\tau$ \cite{Ren2010,Gallavotti1995}:
\begin{equation}
\label{fluc thm}
\lim_{\tau \rightarrow \infty} \frac{1}{\tau}\ln \Bigg [\frac{P_\tau(Q)}{P_\tau(-Q)} \Bigg ] = \frac{Q(\beta_R-\beta_L)}{\tau},
\end{equation}
where $P_\tau(Q)$ denotes the probability distribution of the heat $Q$ transferred from the left to the right bath and $\beta_{L,R}=\frac{1}{k_BT_{L,R}}$. This statement describes the probability at which heat transfer occurs against the thermal gradient. However, in the case of cyclic time-dependent manipulations of the system, it has been shown \cite{Ren2010} that geometric contributions to the heat transfer statistics lead to the addition of correction terms to this theorem. The formalism presented in Sec.~\ref{sec:FCS} allows us to compute the corrections to the fluctuation theorem for  systems in which this geometric term is topologically protected against fluctuations in the driving. 

In order to compute these corrections we start by noticing 
 that the GC fluctuation theorem holds if and only if the characteristic function obeys the Gallavotti-Cohen symmetry $\chi(\lambda) = \chi(-\lambda + i \beta^*)$, where $\beta^* \equiv (\beta_R  - \beta_L) = 0$ for our system of interest, since the temperature of the external leads are assumed to be equal and remain constant throughout the braiding process.  Using this expected symmetry of the CF, we can define the following function quantifying the corrections to the fluctuation theorem:
\begin{eqnarray}
A(\lambda) = |\chi(\lambda) - \chi(-\lambda)|.
\end{eqnarray} 
A non-zero value of $A(\lambda)$ at any value of the counting field, $\lambda$, indicates a correction to the GCFT. 

For the case of small amplitude oscillations in the parameter space of the Majorana Y-junction, we can access the probability distribution for pumped charge via the inverse Fourier transform of the exponentiated total GCF $G(\lambda)$ defined in Eqs.\ \ref{CGF}. The probability distributions for one such cycle are illustrated in Fig. \ref{fig:Prob Small Cycle} for several values of the coupling to the external leads $\Gamma$. Here the asymmetry of the probability distribution with respect to $Q=0$ is clearly visible and corresponds to fact that heat is driven between the external leads, despite the absence of any temperature or chemical potential bias. The inset of Fig. \ref{fig:Prob Small Cycle} illustrates the behaviour of our violation quantifier $A(\lambda)$, which is non-zero and hence indicative of a correction to the FT. The magnitude of the correction also appears to be increasing with $\Gamma$, indicated by the larger area under the graph of $A(\lambda)$. This is a consequence of the increasing translation of $P(Q)$ as an increasing heat current is pumped between the external leads. However, this trend is not general as increasing noise will act to obscure any translation of $P(Q)$ hence reducing the magnitude of the correction function $A(\lambda)$. At the low temperature, relative to the energy scales associated with the scattering matrix, the variance of $P(Q)$ is found to be decreasing with increasing coupling strength (cf.\ Fig.\ \ref{fig:elas heat noise}(a) inset). However, in the high temperature limit, the static noise becomes linearly dependent on $\Gamma$ and we would see that the resulting violation would become less prominent with increasing coupling to the external leads. 
%%%%%%%%%%%%%%%%%%%%%%%%%%%%%%%%%%%%%%
\subsection{Impact upon fluctuation theorems for arbitrary amplitude pumps}
\begin{figure}
	\centering
	\includegraphics[width=0.4\textwidth]{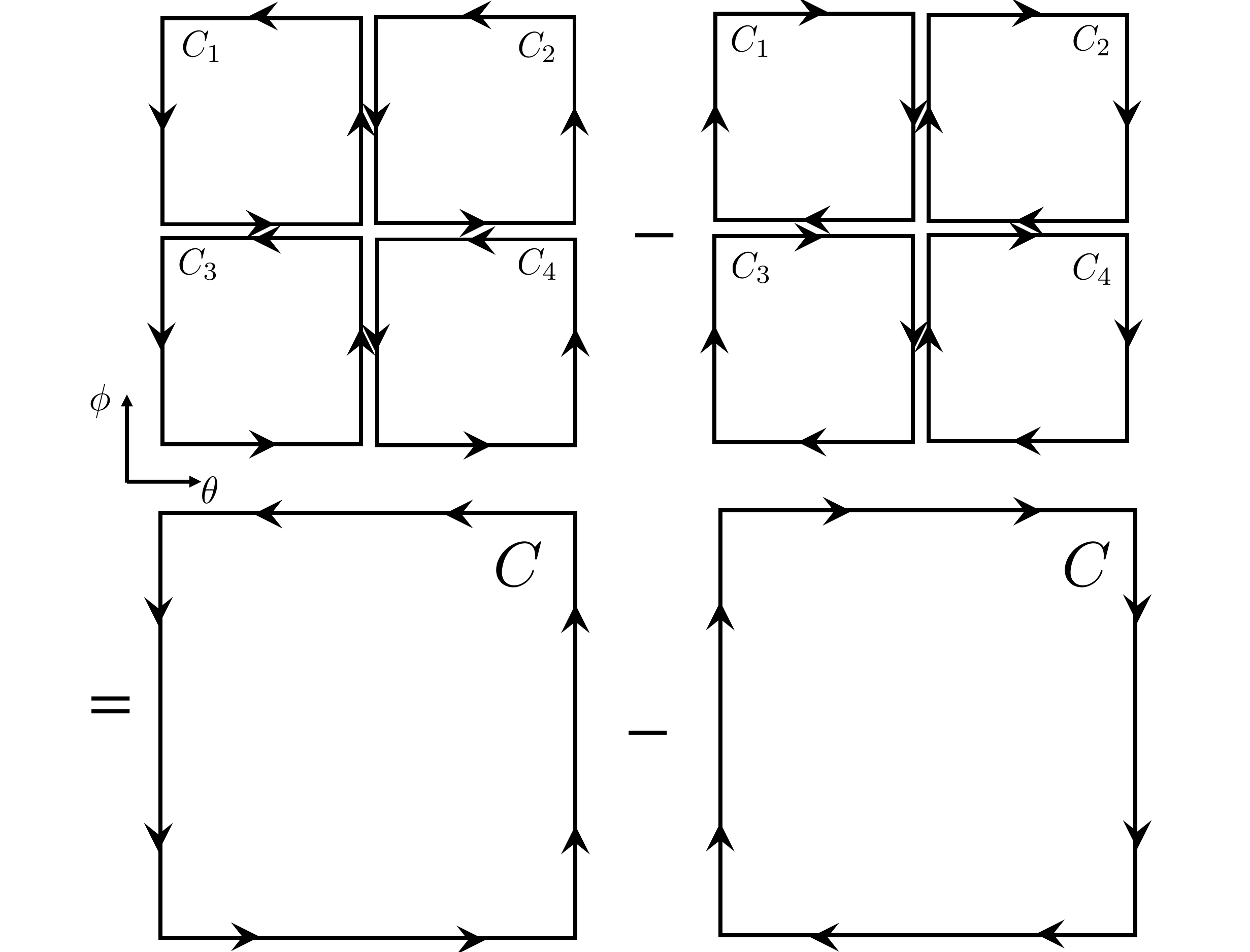} 
	\caption{An illustration of how the difference between contour integrals in opposite directions for arbitrary amplitude cycles, can be broken down into the sum of similar differences on smaller cycles. This result is due to the cancellation of the integrals along the interior sides of the smaller cycles and is valid upon division of the contour $C$ into an arbitrary number of smaller cycles $\{C_i\}$.}
	\label{Contours}
\end{figure}
\label{Large Amp Drive}
For the case of the Majorana braiding [cf.\ Fig.\ \ref{Setup}(b)], the amplitude of the parameter modulation is not small enough for the linear expansion of the scattering matrix in Eq. \ref{S expand} to apply. Yet, in this section, we show that the our results can be extended to demonstrate the correction to the Gallavotti-Cohen fluctuation theorem for large amplitude cycles such as this. Firstly, we make use of the fact that we can write the total generating function for a small amplitude cycle, $G(\lambda)$, as the sum of geometric and dynamic contributions: $G(\lambda) = G^{\mathrm{geom}}(\lambda) + G^{\mathrm{dyn}}(\lambda)$. From our numerical results, we see that for the dynamic contribution, the GC symmetry holds for all $\lambda$ and consequently any corrections to the FT arise solely from the geometric contribution. Hence, an equivalent indicator of FT correction is given by $A^{\mathrm{geom}} = |\chi^{\mathrm{geom}}(\lambda) - \chi^{\mathrm{geom}}(-\lambda)|$, where $\chi^{\mathrm{geom}}(\lambda) = e^{G^{\mathrm{geom}}(\lambda)}$. This quantity, unlike $A(\lambda)$, can be calculated for arbitrary amplitude cycles.

In order to demonstrate this, we define the generating function to be dependent upon the direction we travel around the driving contour in parameter space as $G^\circlearrowleft(\lambda)$ and $G^\circlearrowright(\lambda)$. The difference between these two generating functions $D_Q(\lambda) =G_Q^\circlearrowleft(\lambda)-G_Q^\circlearrowright(\lambda)$ will clearly change sign upon reversal of the pumping direction. For this reason, we can calculate this quantity for a large amplitude pump by dividing the area enclosed by the contour, traversed throughout the braiding process, into smaller areas, within which the weak amplitude approximation is valid. An example of this reasoning is illustrated in Fig. \ref{Contours}. If we write the directional generating functions for each small cycle as closed contour integrals in parameter space, $\ointclockwise _C ds \frac{dt}{ds} G_{\theta,\phi}(\lambda)$, we see that the subtraction of these integrals in opposite directions leads to the cancellation of the interior contributions. This leaves only the desired line integral around the boundary of the larger cycle:
\begin{equation}
\begin{aligned}
D_Q(\lambda) =& \int_{0}^{\mathcal{T}} dt \big( G_{t}^\circlearrowleft(\lambda)-G^\circlearrowright(\lambda,t) \big)
\\
=& \ointclockwise _C ds \frac{dt}{ds} G_{\theta,\phi}(\lambda) - \ointctrclockwise _C ds \frac{dt}{ds} G_{\theta,\phi}(\lambda)
\\
=& \sum_i \Bigg[ \ointclockwise_{C_i} ds_i\frac{dt}{ds_i}  G_{\theta,\phi}(\lambda) - \ointctrclockwise_{C_i} ds_i\frac{dt}{ds_i}  G_{\theta,\phi}(\lambda)\Bigg].
\end{aligned}
\end{equation}
We can hence obtain the quantity $D_Q(\lambda)$ for large amplitude cycles by summing the contributions of cycles in which the small amplitude approximation is valid. In the limit that the interior cycles are made infinitesimally small, the summation of the contributions from each smaller cycle can be used to approximate the form of the generating function resulting from the traversal of a contour of an arbitrary shape and size.

The quantity $D_Q(\lambda)$ isolates the contribution to the generating function which is sensitive to the direction in which the pumping contour is traversed. One can show that, for a small amplitude, two parameter pump, this term is the proportional to $X_{\omega,1}X_{\omega,2}$ and corresponds to $2G^{\mathrm{geom}}(\lambda)$. We can hence access the violation function $A^{\mathrm{geom}}(\lambda)$ for arbitrary amplitude cycles. 
\begin{figure}
	\centering
	\includegraphics[width=0.45\textwidth]{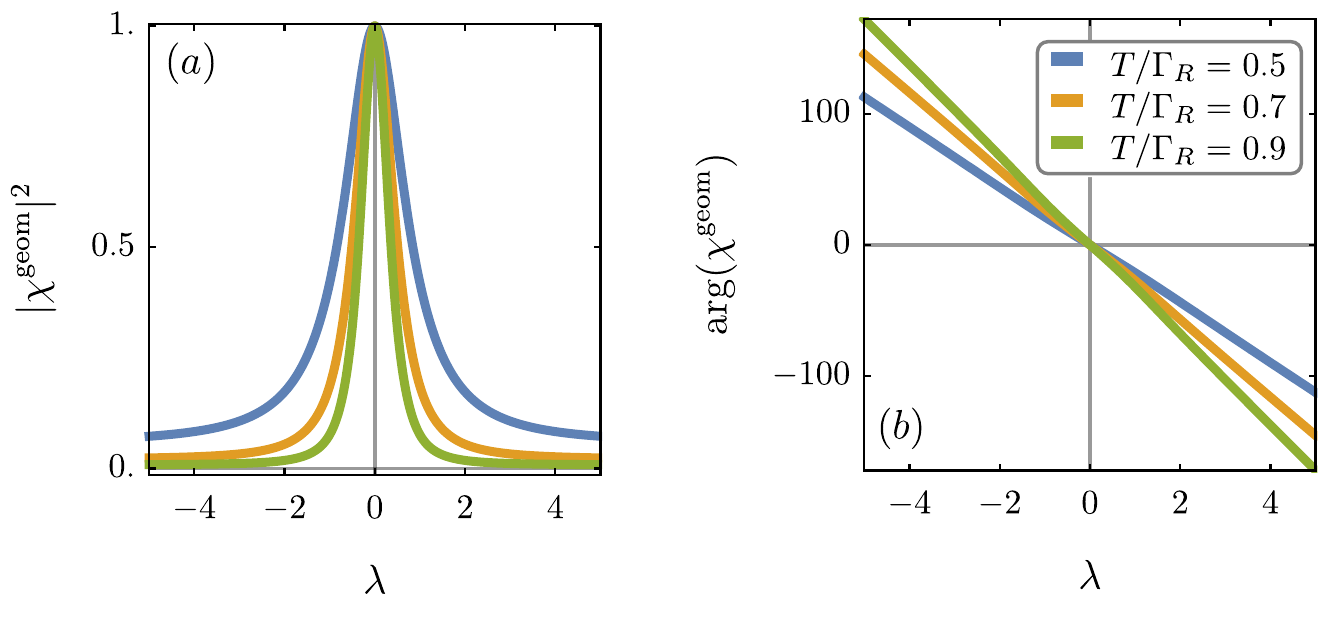}
	\caption{Absolute value (a) and argument (b) of the geometric contribution to the heat transport characteristic function $\chi^{\mathrm{geom}}(\lambda)$ for the case of a Majorana braiding protocol. Results are plotted for several values of the external lead temperature. Asymmetry of this function in $\lambda$ indicates an apparent violation of the Gallavotti-Cohen type fluctuation theorem.}
	\label{GF}
\end{figure}
%The Gallavotti-Cohen fluctuation theorem can be formulated regardless of the direction of the pumping cycle. In the case that both of the external leads are held at the same temperature this gives us that
%\begin{equation}
%\label{directional FT}
%\frac{P^\circlearrowleft(Q)}{P^\circlearrowleft(-Q)} = %\frac{P^\circlearrowright(Q)}{P^\circlearrowright(-Q)} = 1.
%\end{equation}
%When this relationship holds, the Gallavotti-Cohen symmetry should be found to be present in both of the directional generating functions so that $G_Q^{\circlearrowleft(\circlearrowright)} (\lambda) = G_Q^{\circlearrowleft(\circlearrowright)} (-\lambda)$. As a result this symmetry should also be present in any linear combination of this quantities and in particular we would expect that $D_Q(\lambda) = D_Q(-\lambda)$.

The absolute value and argument of $\chi^{\mathrm{geom}}(\lambda)$ are plotted in Fig.\ \ref{GF}(a,b) for a Majorana braiding process. It can be seen that, although the GC symmetry is present in the real part of this quantity, the imaginary part is non-zero and antisymmetric with respect to $\lambda$. As a result, the violation function takes the form
\begin{equation}
A^{\mathrm{geom}}(\lambda)=2|\chi^{\mathrm{geom}}| |\sin\Big(\arg(\chi^{\mathrm{geom}})\Big)|
\end{equation}  
and corrections are clearly required to the GCFT when considering a Majorana braiding process. Furthermore, the presence of this correction does not require the modulation of the external temperature gradient as in topologically trivial systems \cite{Ren2010} and hence stems solely from the cyclic variation of the systems internal parameters. 

In order to consider the temperature dependence of this apparent violation one must take into account two competing factors.  Although the pumped heat increases as a function of $T$, illustrated by the increasing gradient of $\arg(\chi^{\mathrm{geom}})(\lambda)$ in Fig.\ \ref{GF}(b), we also know that the second cumulant of the pumped heat, $\mathcal{M}^{(2)}$ varies as $T^5$. This increased variance, indicated by the rate of decay of $|\chi^{\mathrm{geom}}|^2$ plotted in Fig.\ \ref{GF}(a), leads to the overlap of the probability distributions $P(Q)$ and $P(-Q)$ and hence a reduction in the correction to the GCFT. Furthermore, the fact that the correction is purely geometric in nature, means that, for $T\gg\omega$, the correction is dependent only on the contour traversed in parameter space and is independent of the driving frequency. Given that, in the case of a Majorana braiding, this contour is topologically protected against fluctuations in the driving, the behaviour of the violation function $A^{\mathrm{geom}}(\lambda)$ will also exhibit this protection.   
\color{black}
%%%%%%%%%%%%%%%%%%%%%%%%%%%%%%%%%%%%%%%%%%%%%%%%%%%%%
%%%%%%%%%%%%%%%%%%%%%%%%%%%%%%%%%%%%%%%%%%%%%%%%%%%%%%
\section{Conclusions}

A system driven in an adiabatic cycle shows corrections to thermodynamic fluctuation theorems which depend on geometric properties of the cycle, as opposed to its dynamical features.
Here we have studied the statistics of heat transfer for adiabatic cycles associated with the topologically protected evolution of a quantum system, specifically, a 1-dimensional topological superconductor undergoing braiding of its Majorana zero modes. We have first obtained general expressions for the statistics of heat transfer, which extend known results for the charge transport full counting statistics. We singled out the peculiarities of Majorana zero modes in the heat and charge current noise, including a correction to the Gallavotti-Cohen type fluctuation theorem. We have successfully extended this result to finite amplitude cycles and showed that the heat transfer associated with Majorana braiding induces a correction to a Gallavotti-Cohen type fluctuation theorem. As opposed to analogous corrections in non-topological systems which require cyclical variation of the external temperatures \cite{Ren2010}, our contribution stems solely form a cycle in the system's parameter space at constant temperature gradient and is a result of the coherent dynamics of the driving. Moreover, the correction term is geometric in nature and topologically protected against small, slow fluctuations of the driving.
The identification of corrections to transport fluctuation theorems, in terms of quantum coherent contributions to scattering processes, allows for further investigation to incorporate such contributions in properly modified fluctuation theorems.

\section{Acknowledgments}
A.R. acknowledges the UK  Engineering  and Physical Sciences Research Council (EPSRC) via Grant No.  EP/P010180/1. D.M. acknowledges support from the Israel Science Foundation (Grant No. 1884/18).

%%%%%%%%%%%%%%%%%%%%%%%%%%%%%%%%%%%%%%%%%%%%%%%%%%%%%

%%%%%%%%%%%%%%%%%%%%%%%%%%%%%%%%%%%%%%%%%%%%%%%%%%%%%%

%%%%%%%%%%%%%%%%%%%%%%%%%%%%%%%%%%%%%%%%%%%%%%%%%%%%%
\begin{widetext}
\appendix
\section{Characteristic Function for Driven Heat Transport}
\label{CF derivation}
In order to define the characteristic function for the case of the Majorana braiding setup, which will include inelastic scattering between the nearest energy sidebands, we must consider both the time and energy dependence of the scattering matrix. Firstly, in order to deal with the time dependence, we will divide the total time interval $t_0$ into discrete time steps, which we will denote as $t_i$, before taking the continuum limit. At each time step the components of the scattering matrix will be assumed constant and we assume that the outgoing distribution of particles at a time $t_i$ has no influence upon the ingoing distribution at a later time, since this is always taken as the equilibrium Fermi distribution of the particle bath. As a consequence, the probabilities of the heat transferred across the junction within each time interval can be taken as independent from one another. We can write the characteristic function associated with a heat transfer $Q$ after a total time $t_0$ as 
\begin{equation}
\chi_Q(\lambda) = \sum_Q e^{i\lambda Q} \sum_{\{q_{t_i}\}} \delta \Big (Q=\sum_{t_i}q_{t_i} \Big ) P(q_{t_1},q_{t_2},...),
\end{equation}
where $\{q_{t_i}\}$ denotes all possible combinations of the heat quantities $q_{t_i}$. The independence of the probability distribution at each time allows this expression to be rewritten as the product of characteristic functions defined at each discrete time step:
\begin{equation}
\begin{aligned}
\chi_Q(\lambda) =& \sum_Q e^{i\lambda Q} \sum_{\{q_{t_i}\}} \delta \Big (Q=\sum_{t_i}q_{t_i} \Big ) \prod_{t_i} P(q_{t_i}) 
\\
=&  \sum_{\{q_{t_i}\}}  e^{i\lambda \sum_{t_i} q_{t_i}} \prod_{t_i} P(q_{t_i}) 
\\
=& \sum_{q_{t_1}} e^{i\lambda  q_{t_1}} \sum_{q_{t_2}}e^{i\lambda  q_{t_2}} ... \prod_{t_i} P(q_{t_i})  
\\
=& \prod_{t_i} \sum_{q_{t_i}} e^{i\lambda  q_{t_i}}P(q_{t_i})  = \prod_{t_i} \chi_{t_i}(\lambda).
\end{aligned}
\end{equation}
Taking the logarithm of this expression we can define the characteristic function the time continuum limit:
\begin{equation}
\ln(\chi_Q(\lambda)) = \int_{0}^{t_0} dt \ln(\chi_{t}(\lambda)).
\end{equation}

It now remains to find an appropriate expression for the function $\chi_{t_i}(\lambda)$. The scattering between nearest energy sidebands means that we cannot consider the characteristic function at each energy independently, a fact which complicates the subsequent calculation significantly. With this in mind, we can define the net number of particles with energy $\epsilon_j$ traveling to the right in the left lead as $q_{t_i}(\epsilon_j) = m_{t_i}(\epsilon_j) - n_{t_i}(\epsilon_j)$ and thus write the characteristic function at a fixed time $t_i$ as
\begin{equation}
\label{Characteristic fnc}
\begin{aligned}
\chi_{t_i}(\lambda) =& \sum_{q_{t_i}} e^{i\lambda q_{t_{i}}} \sum_{\{q_{t_i}(\epsilon_j)\}} \delta \Big (q_{t_i} = \sum_{\epsilon_j} \epsilon_j q_{t_i}(\epsilon_j) \Big ) P(q_{t_i}(\epsilon_1),q_{t_i}(\epsilon_2),...)
\\
=& \sum_{\{q_{t_i}(\epsilon_j)\}} e^{i\lambda \sum_{\epsilon_j} \epsilon_j q_{t_i}(\epsilon_j)} P(q_{t_i}(\epsilon_1),q_{t_i}(\epsilon_2),...)
\\
=&
\sum_{\{m_{t_i}(\epsilon_j),n_{t_i}(\epsilon_j)\}} e^{i\lambda \sum_{\epsilon_j} \epsilon_j (m_{t_i}(\epsilon_j)-n_{t_i}(\epsilon_j))}   P(m_{t_i}(\epsilon_1),-n_{t_i}(\epsilon_1),...)
\\
=&  \Big < e^{i\lambda \hat{Q}_\rightarrow}e^{-i\lambda \hat{Q}_\leftarrow} \Big >,
\\ 
\end{aligned}
\end{equation}
where
\begin{equation}
\label{heat number}
\hat{Q}_{\rightarrow(\leftarrow)} = \sum_{\epsilon_i} \epsilon_i \Big (\hat{N}^e_{\epsilon_i \rightarrow(\leftarrow)}+\hat{N}^h_{\epsilon_i \rightarrow(\leftarrow)} \Big ).
\end{equation}

%%%%%%%%%%%%%%%%%%%%%%%%%%%%%%%%%%%%%%%%%%%%%%%%%%%%%%
%%%%%%%%%%%%%%%%%%%%%%%%%%%%%%%%%%%%%%%%%%%%%%%%%%%%%%

\section{Projective nature of Number Operator Matrices}
\label{projector derivation}
Here we will show how products of number operators, defined in Sec.\ \ref{sec:Adiabatic}, which act at the same or different energies can be simplified in order to demonstrate their projective nature. In order to do this we must first consider several relationships that can be obtained from the unitarity of the scattering matrix. Treating the discretised particle energy levels as ingoing and outgoing propagation channels, we can write the scattering matrix in block form,
\begin{equation}
\hat{S} = \begin{pmatrix}
\ddots & & & & & \bigzero &
\\
& S^{\alpha \beta}_\omega(\epsilon_i) & S^{\alpha \beta}(\epsilon_i) & S^{\alpha \beta}_{-\omega}(\epsilon_i) & &
\\
& & S^{\alpha \beta}_\omega(\epsilon_{i+1}) &  S^{\alpha \beta}(\epsilon_{i+1}) &  S^{\alpha \beta}_{-\omega}(\epsilon_{i+1}) &
\\
& & & S^{\alpha \beta}_\omega(\epsilon_{i+2}) &  S^{\alpha \beta}(\epsilon_{i+2}) &  S^{\alpha \beta}_{-\omega}(\epsilon_{i+2}) 
\\
&  \bigzero & & & &  \ddots
\end{pmatrix}•.
\end{equation}•
By writing the scattering matrix in this form we can immediately see that its unitarity gives us the following useful relations:
\begin{subequations}
	\begin{align}
	\label{relation 1}
	|S^{\alpha \beta}(\epsilon_i)|^2 + |S^{\alpha \beta}_\omega(\epsilon_i)|^2  +|S^{\alpha \beta}_{-\omega}(\epsilon_i)|^2  = \mathcal{I},
	\\
	\label{relation 2}
	S^{\alpha \beta}(\epsilon_i){S^{\beta \gamma}_\omega(\epsilon_{i+1})}^\dagger + S^{\alpha \beta}_{-\omega}(\epsilon_i){S^{\beta \gamma}(\epsilon_{i+1})}^\dagger  = 0,
	\\
	\label{relation 3}
	S^{\alpha \beta}_{-\omega}(\epsilon_i){S^{\beta \gamma}_\omega(\epsilon_{i+2})}^\dagger = 0.
	\end{align}
\end{subequations}•
These relationships can be further simplified by considering the symmetric nature of each of he blocks, so that ${S^{\alpha \beta}(\epsilon_i)}^\dagger =  {S^{\alpha \beta}(\epsilon_i)}^*$ and  ${S^{\alpha \beta}_{\pm \omega}(\epsilon_i)}^\dagger =  {S^{\alpha \beta}_{\pm \omega}(\epsilon_i)}^*$. 

Now let us consider for example the square of the outgoing number operator matrix at energy $\epsilon_i$:
\begin{equation}
\nonumber
\begin{aligned}
P_{\epsilon_i \leftarrow}^2 = (P^e_{\epsilon_i \leftarrow} + P^h_{\epsilon_i \leftarrow})^2 = {P^e_{\epsilon_i \leftarrow}}^2 +  {P^h_{\epsilon_i \leftarrow}}^2 + P^e_{\epsilon_i \leftarrow}P^h_{\epsilon_i \leftarrow} + P^h_{\epsilon_i \leftarrow} P^e_{\epsilon_i \leftarrow}.
\end{aligned}•
\end{equation}•  
First considering the top left non-zero element of the matrix for the squared electron term, we have that
\begin{equation}
\begin{aligned}
\{{P^e_{\epsilon_i \leftarrow}}^2\}_{ii} =&  {S_\omega^{\alpha 1}(\epsilon_i)}^*S_\omega^{1 \beta}(\epsilon_i)  {S_\omega^{\beta 1}(\epsilon_i)}^*S_\omega^{1 \gamma}(\epsilon_i) + 
{S_\omega^{\alpha 1}(\epsilon_i)}^*S^{1 \beta}(\epsilon_i)  {S^{\beta 1}(\epsilon_i)}^*S_\omega^{1 \gamma}(\epsilon_i)  \\ + &
{S_\omega^{\alpha 1}(\epsilon_i)}^*S_{-\omega}^{1 \beta}(\epsilon_i) {S_{-\omega}^{\beta 1}(\epsilon_i)}^*S_\omega^{1 \gamma}(\epsilon_i)
\\
=&  {S_\omega^{\alpha 1}(\epsilon_i)}^* \Big (S_\omega^{1 \beta}(\epsilon_i)  {S_\omega^{\beta 1}(\epsilon_i)}^* + S^{1 \beta}(\epsilon_i)  {S^{\beta 1}(\epsilon_i)}^* +  S_{-\omega}^{1 \beta}(\epsilon_i) {S_{-\omega}^{\beta 1}(\epsilon_i)}^* \Big ) S_\omega^{1 \gamma}(\epsilon_i)
\\
=&  {S_\omega^{\alpha 1}(\epsilon_i)}^* S_\omega^{1 \gamma}(\epsilon_i) = \{P^e_{\epsilon_i \leftarrow}\}_{ii},
\end{aligned}
\end{equation}
where in the final step we have made use of the unitarity relation given in Eq. \ref{relation 1}. The same reasoning can be applied to the other components of the matrix and in the case of the outgoing number operator for holes. Hence we can conclude that $  {P^e_{\epsilon_i \leftarrow}}^2 = P^e_{\epsilon_i \leftarrow}$ and $  {P^h_{\epsilon_i \leftarrow}}^2 = P^h_{\epsilon_i \leftarrow}$. Next considering the cross-terms in the expansion we see that,
\begin{equation}
\begin{aligned}
\{ P^e_{\epsilon_i \leftarrow}P^h_{\epsilon_i \leftarrow}\}_{ii} =&  {S_\omega^{\alpha 1}(\epsilon_i)}^*S_\omega^{1 \beta}(\epsilon_i)  {S_\omega^{\beta 2}(\epsilon_i)}^*S_\omega^{2 \gamma}(\epsilon_i) + 
{S_\omega^{\alpha 1}(\epsilon_i)}^*S^{1 \beta}(\epsilon_i)  {S^{\beta 2}(\epsilon_i)}^*S_\omega^{2 \gamma}(\epsilon_i)  \\ + &
{S_\omega^{\alpha 1}(\epsilon_i)}^*S_{-\omega}^{1 \beta}(\epsilon_i) {S_{-\omega}^{\beta 2}(\epsilon_i)}^*S_\omega^{2 \gamma}(\epsilon_i)
\\
=&  {S_\omega^{\alpha 1}(\epsilon_i)}^* \Big (S_\omega^{1 \beta}(\epsilon_i)  {S_\omega^{\beta 2}(\epsilon_i)}^* + S^{1 \beta}(\epsilon_i)  {S^{\beta 2}(\epsilon_i)}^* +  S_{-\omega}^{1 \beta}(\epsilon_i) {S_{-\omega}^{\beta 2}(\epsilon_i)}^* \Big ) S_\omega^{2 \gamma}(\epsilon_i)
\\
=& 0,
\end{aligned}
\end{equation}
where once again we have used Eq. \ref{relation 1}. This results hold for all elements of the matrices $ P^e_{\epsilon_i \leftarrow}P^h_{\epsilon_i \leftarrow}$ and $P^h_{\epsilon_i \leftarrow} P^e_{\epsilon_i \leftarrow}$ and hence we have demonstrated the projective nature of the outgoing number operator matrices at each energy: $P_{\epsilon_i \leftarrow}^2 = P_{\epsilon_i \leftarrow}$.

Next we will show that, in addition to this result, the sum of the matrices $P_{\epsilon_i \leftarrow}$ over all energies is also itself a projector. In order to do this we will evaluate the product 
\begin{equation}
\nonumber
\Big ( \sum_i  P^e_{\epsilon_i \leftarrow}+P^h_{\epsilon_i \leftarrow} \Big )^2.
\end{equation}•
Due to the shape of the $P$ matrices given in Eq. \ref{outgoing P}, we find that the only non-zero contributions to this product take the form:
\begin{equation}
\begin{aligned}
\Big  (P^e_{\epsilon_i \leftarrow} + P^h_{\epsilon_i \leftarrow} \Big )\Big  (P^e_{\epsilon_j \leftarrow} + P^h_{\epsilon_j \leftarrow}\Big ) \ \mathrm{where} \ |i-j| \leq 2.
\end{aligned}
\end{equation}
For the case in which $i=j$ we have already shown that these matrices are projectors. Next considering the case $|i-j|=1$, we will first consider the elements of the matrix $P^e_{\epsilon_i \leftarrow} P^e_{\epsilon_{i\pm1}\leftarrow}$. In particular the top left non-zero element of this matrix will be of the form
\begin{eqnarray}
\begin{aligned}
\{ P^e_{\epsilon_i \leftarrow} P^e_{\epsilon_{i+1}\leftarrow} \}_{i-1 i} =& {S_\omega^{\alpha 1}(\epsilon_i)}^*S^{1 \beta}(\epsilon_{i}) {S_\omega^{\beta 1}(\epsilon_{i+1})}^*S_\omega^{1 \gamma}(\epsilon_{i+1}) + {S_\omega^{\alpha 1}(\epsilon_i)}^*S_{-\omega}^{1 \beta}(\epsilon_i)  {S^{\beta 1}(\epsilon_{i+1})}^*S_\omega^{1 \gamma}(\epsilon_{i+1})
\\
=& {S^{\alpha 1}(\epsilon_i)}^* \Big (S^{1 \beta}(\epsilon_{i}) {S_\omega^{\beta 1}(\epsilon_{i+1})}^* + S_{-\omega}^{1 \beta}(\epsilon_i)  {S^{\beta 1}(\epsilon_{i+1})}^*  \Big ) S_\omega^{1 \gamma}(\epsilon_{i+1})
\\
=& 0.
\\
\{ P^e_{\epsilon_i \leftarrow} P^e_{\epsilon_{i-1}\leftarrow} \}_{i-1 i} =& {S_\omega^{\alpha 1}(\epsilon_i)}^*S^{1 \beta}_\omega(\epsilon_{i}) {S^{\beta 1}(\epsilon_{i-1})}^*S_\omega^{1 \gamma}(\epsilon_{i-1}) + {S_\omega^{\alpha 1}(\epsilon_i)}^*S^{1 \beta}(\epsilon_i)  {S_{-\omega}^{\beta 1}(\epsilon_{i-1})}^*S_\omega^{1 \gamma}(\epsilon_{i-1})
\\
=& {S_\omega^{\alpha 1}(\epsilon_i)}^* \Big (S_\omega^{1 \beta}(\epsilon_{i}) {S^{\beta 1}(\epsilon_{i-1})}^* + S^{1 \beta}(\epsilon_i)  {S_{-\omega}^{\beta 1}(\epsilon_{i-1})}^*  \Big ) S_\omega^{1 \gamma}(\epsilon_{i-1})
\\
=& 0.
\end{aligned}
\end{eqnarray}
Here we have used the relation given in Eq. \ref{relation 2} and this relationship can be shown to hold true for every element of this matrix. In the case of holes, we also have that,
\begin{eqnarray}
\begin{aligned}
\{ P^h_{\epsilon_i \leftarrow} P^h_{\epsilon_{i+1}\leftarrow} \}_{i-1 i} =& {S_\omega^{\alpha 1}(\epsilon_i)}^*S^{1 \beta}(\epsilon_{i}) {S_\omega^{\beta 1}(\epsilon_{i+1})}^*S_\omega^{1 \gamma}(\epsilon_{i+1}) + {S_\omega^{\alpha 1}(\epsilon_i)}^*S_{-\omega}^{1 \beta}(\epsilon_i)  {S^{\beta 1}(\epsilon_{i+1})}^*S_\omega^{1 \gamma}(\epsilon_{i+1})
\\
=& {S^{\beta 1}(\epsilon_i)}^*S_\omega^{1 \gamma}(\epsilon_i) \Big (S^{1 \beta}(\epsilon_{i}) {S_\omega^{\beta 1}(\epsilon_{i+1})}^* + S_{-\omega}^{1 \beta}(\epsilon_i)  {S^{\beta 1}(\epsilon_{i+1})}^*  \Big ) S_\omega^{1 \gamma}(\epsilon_{i+1})
\\
=& 0.
\end{aligned}
\end{eqnarray}
In the same way, it can be shown that the product terms between electrons and holes are also zero so that $\Big  (P^e_{\epsilon_i \leftarrow} + P^h_{\epsilon_i \leftarrow} \Big )\Big  (P^e_{\epsilon_j \leftarrow} + P^h_{\epsilon_j \leftarrow}\Big )=0 $ when $|i-j|=1$. Finally for the case of $|i-j|=2$, we will again consider as an example the top left non-zero element of the relevant matrix:
\begin{eqnarray}
\begin{aligned}
\{ P^e_{\epsilon_i \leftarrow} P^e_{\epsilon_{i+1}\leftarrow} \}_{i-1 i+1} =& {S_\omega^{\alpha 1}(\epsilon_i)}^*S_{-\omega}^{1 \beta}(\epsilon_{i}) {S_\omega^{\beta 1}(\epsilon_{i+2})}^*S_\omega^{1 \gamma}(\epsilon_{i+2})  
\\
=& 0,
\end{aligned}
\end{eqnarray}
by Eq. \ref{relation 3}. We have hence demonstrated the projective nature of the sum of the number operator matrices,
\begin{equation}
\Big ( \sum_i  P^e_{\epsilon_i \leftarrow}+P^h_{\epsilon_i \leftarrow} \Big )^2 = \sum_i  {P^e_{\epsilon_i \leftarrow}}^2+{P^h_{\epsilon_i \leftarrow}}^2 =  \sum_i  P^e_{\epsilon_i \leftarrow}+P^h_{\epsilon_i \leftarrow}.
\end{equation}
\end{widetext}

\bibliography{FCSPumping}  
\end{document}